\documentclass[%11pt,twocolumn,twoside,
a4paper,amsmath,amssymb,aps,showkeys,showpacs]{revtex4}

\usepackage[cp1251]{inputenc}
\usepackage{multirow}

\usepackage{graphics} %   <------- for LatTeX + PS
\usepackage{epstopdf}
\usepackage{epsfig}
\usepackage{amsfonts,amssymb,latexsym}
\usepackage{amsmath}
\usepackage{amsthm}
\usepackage{amstext}
\renewcommand{\theequation}{\thesection.\theequation}
\numberwithin{equation}{section}

\begin{document}
\DeclareGraphicsExtensions{.jpg,.pdf,.mps,.png} %   <------- for PDFLatTeX

%*****************   The Body of the Article:   *************************

%\rhead[]{}%<------
%\lhead[]{}%<------
\title{Vortex dynamics of charge carriers in the quasirelativistic
graphene model : high-energy $\vec k\cdot \vec p$ approximation }

\author{Halina~V.~Grushevskaya}
\email{grushevskaja@bsu.by} \affiliation{Physics Department,
Belarusian State University, 4 Nezalezhnasti Ave., 220030 Minsk,
BELARUS}
\author{George~Krylov}
\email{krylov@bsu.by} \affiliation{Physics Department, Belarusian
State University, 4 Nezalezhnasti Ave., 220030 Minsk, BELARUS}

\received{08 November, 2019}



\begin{abstract}
 Within the earlier developed high-energy-$\vec k\cdot \vec p$-Hamiltonian  approach to describe graphene-like  materials the
simulations of non-Abelian Zak phases and band structure of the
quasi-relativistic graphene  model with a flavors number $N=3$
have been performed in approximations with %of zero- and non-zero values of
and without gauge  fields (flavors).
%The approximation of non-zero gauge fields turns out topologically non-trivial one.
%The approximation of zero gauge fields is topologically trivial one due to hexagonal symmetry.
It has been shown that a Zak-phases set for non-Abelian
Majorana-like excitations (modes) in Dirac valleys
 of the quasirelativistic graphene model
is the cyclic group $\mathbf{Z}_{12}$. This group is deformed into
$\mathbf{Z}_8$ at sufficiently high momenta due to  deconfinement
of the modes.    Since the deconfinement removes the degeneracy of
the eightfolding valleys, Weyl nodes  and antinodes emerge.
We offer that a Majorana-like  mass term of the quasirelativistic
model effects  on  the graphene band structure in a following way.
Firstly  the inverse symmetry  emerges at "switching on"\ of the
mass term, and  secondly the mass term shifts the location of Weyl
nodes and antinodes into the region of higher energies and
accordingly the  Majorana-like modes can exist without mixing with
the nodes.
\end{abstract}

 \maketitle

 \maketitle

\section{Introduction}

%В настоящее время топологические графеноподобные материалы
%привлекают огромное внимание из-за возможности реализовать
% устойчивые квантовые вычисления на квантовых устройствах, сконструированных из таких материалов.
 %
Currently, topological graphene-like materials attract great
attention due to  possibility of implementing robust quantum
computing on quantum devices constructed from such materials.
%
%Типы
%топологических дефектов зонной структуры графеноподобных
%материалов разнообразны:  Weyl
%nodes and antinodes, Weyl nodes (antinodes) line, ????drumhead %барабанные
%состояния, flat bands, zero Majorana modes,  несмотря на то, что
%кристаллическая структура всех топологических материалов или
%гесагональна, или почти гексагональна [???].
%
Types of topological defects in the band structure of
graphene-like materials are diverse: Weyl nodes and antinodes \cite{Lu-Shen2017Front-Phys},
Weyl nodal (antinodal) lines \cite{NatCommunic7-2016Bian},
drumhead-like surface flat bands \cite{PhysRevX6-2016Muechler,Honeycomb-KagomeLattice2017}, zero-energy
Majorana modes  \cite{PhysRevX5-2015San-Jose}, despite the fact that the crystal structure of all
topological  materials is either hexagonal or almost
hexagonal \cite{SangwanHersam2018}.
%
%Установление механизма влияния степени нарушения симметрии
% by спин-орбитальным взаимодействием на тип дефектности зонной структуры - трудная задача, но экстремально
% важная для приложений топологических материалов.
 %
 Establishing a mechanism of the influence of the degree of symmetry breaking
 due to spin-orbit interaction on the type of  the band-structure defectiveness
 is a challenge, but it is extremely important for applications of topological materials.
 %
%Сложность решения данной проблемы заключается в  невозможности
%сконструировать maximally localized Wannier orbitals
%максимально локализованные орбитали Ванье
%для зонной структуры с топологическими дефектами [].
%
%Difficulty
A hindrance to solving this problem lies in impossibility to
construct maximally localized Wannier orbitals in a lattice site $i$ for a band
structure with topological defects owing to the presence
%наличия
of the defect in the site $i$. %[] .
 Majorana end states were implemented  as
subgap levels of an atomic chain  on the surface
%на поверхности
of a p-wave
superconductor. This system is named  %называется
 a Kitaev's chain %!!!!!!!!!!!!!!!!!!!!!!
\cite{Physics-Uspekhi44-2001Kitaev,JPhysB40-2007Semenoff}.
%[A. Kitaev.  Unpaired Majorana fermions in quantum wires.  Phys. Usp. {\bf 44}(Suppl.), 131–36 (2001). cond-mat/0010440.
%G.W. Semenoff, P. Sodano. Stretched quantum states emerging from a Majorana medium. J. Physics B. {\bf 40}, 1479 (2007).].
These %Эти
subgap levels are similar to Shockley or Shiba states bound
to electric or magnetic end impurities %!!!!!!!!!!!!!!!!%%%%
(see \cite{%C.W.J.
Beenakker,Peng} %. Search for Majorana Fermions in Superconductors. Annu. Rev. Condens. Matt. Phys.  {\bf 4}, 113–136 (2013).
%[(for Shiba) Ya.Peng, F.Pientka, Yu.Vinkler-Aviv, L.I.Glazman, F. von Oppen.
%Robust Majorana Conductance Peaks for a Superconducting Lead //PRL {\bf 115}, 266804 (2015)
and references therein) %и ссылки там ],
but %но
Shockley or Shiba states do not possess topological stability of vortex states.
%не обладают топологической устойчивостью вихревых состояний.
By connecting three %Соединив три
Kitaev's chains into Y-like form  and tuning interactions
%цепочки Y-образно  и настроив взаимодействие
$\Delta_i$, $i=1,2,3$ of the edge Majorana fermions
in that Y-trijunction, it is possible to force two %можно заставить  два
Majorana midgap states from three ones at the %из трех на
ends %(концах)
of the trijunction alternatively change their positions %меняться местами попеременно !!!!!
\cite{%[ B. van
Heck}. %,  A.R. Akhmerov,  F. Hassler,  M. Burrello,  C.W.J. Beenakker.  New J. Phys. {\bf 14}, 035019 (2012).].
The main problem %Основная проблема
of the Kitaev's-chain network %and the Y-shaped trijunction
is the broadening of  Majorana chain-end states
in the place of contacts between the chains %or between Weyl nodes (antinodes) and Cooper pairs
%-- уширение майорановских состояний в местах контактов цепочек,
and, respectively, the small lifetime of  Majorana quasi-particle excitations in the Kitaev's chain.
 As it turns out %Как оказалось !!!!!!!!!!!!!!!!!!!
\cite{Fu-Kane2008PhysRev},
%[L. Fu, C.L. Kane. Superconducting Proximity Effect and Majorana Fermions
%at the Surface of a Topological Insulator. Phys. Rev. Lett. {\bf 100}, 096407 (2008). DOI: 10.1103/PhysRevLett.100.096407],
an interface between a topological isolator (TI) and a s-wave superconductor can be described by the same
system of equations as for %описывается такой же системой уравнений, что и
the Kitaev's chain.
Zero-energy Weyl nodes and anti-nodes (monopoles) % ноды и антиноды
in TI-surface band structure play a role %играют роль
of Majorana-vortex cores, and Cooper pairs are a "feathering"\ of these %оперением этих
cores. The advantage of the implementation of motionless
Majorane-like excitations on the edge of metallic TI-surfaces is
an opportunity to realize an interchanging (one dimensional
braiding) of Majorana particles among themselves on a contactless
Y-shaped Josephson junction which increases the Majorana lifetime
\cite{IntJModPhys2014Ge-Yu,NayakWilczek,Ivanov,Kitaev2003,Simon,Oppen2015}.
%Преимущество реализации неподвижных майорано-подобных возбуждений на краю of metallic TI-поверхности  заключается в возможности
%реализовать an interchanging (braiding) of майорановских частиц между собой на
%безконтактном Y-образном джозефсоновском (Josephson) соединении that увеличивает майорановское время жизни.
%Motionless Majorane-like excitations in the Y-trijunction
%are driven by the phase $ \phi $ change of the  wave function of Cooper pairs.
%Неподвижные майорано-подобные возбуждения в the trijunction приводятся в движение изменением фазы
%$\phi$ волновой функции куперовских пар.

A chiral massless Dirac fermion (helical state) acquires a phase
% приобретает фазу
$\pi$ when bypassing on a closed loop due
to
%при обходе по замкнутому контуру из-за
its spin (helicity) pointing in the direction of motion. Two Majorana bound states, which form this Dirac fermion
%которые формируют этот
%дираковский фермион,
are interchanged at such bypass by Dirac fermion
% обмениваются местами при обходе дираковского фермиона по замкнутому контуру,
and respectively
one more interchange of the Majorana fermions is necessary to place
them on their original positions.
%необходим еще один обмен майорановскими фермионами, чтобы вернуть их на свои места.
Therefore, at single interchange of Majorana particles
their wave function gains only half  of the phase ($\pi/2$)
%при одном обмене майорановских частиц их волновая функция приобретет в два раза меньшую фазу
 of that gained by the massless Dirac fermion the particles compose.
%,
%чем безмассовый дираковский фермион, который они составляют.
Accordingly, on every Majorana fermion in the pair
the
%на один майорановский фермион в паре приходится
phase shift $\pi/4$ is accounted for
signifying that  a statistics of the Majorana
%майорановские
helical edge states
%состояния
is non-Abelian one.
%The Majorana fermions  are braided mathematically by a non-Abelian unitary
%transformation that a non-Abelian Berry phase changes  on %изменяется на
%$ {\pi/4}$ %!!!!!
%\cite{IntJModPhys2014Ge-Yu,NayakWilczek,Ivanov,Kitaev2003,Simon,Oppen2015}.
%[ C. Nayak,  F. Wilczek.  Nucl. Phys. B {\bf 479}, 529 (1996). D. Ivanov.  Phys. Rev. Lett. {\bf 86}, 268 (2001).
%A.Y. Kitaev.  Ann. Phys. {\bf 303}, 2 (2003). C. Nayak,  S.H. Simon,  A. Stern,  M. Freedman, S. Das Sarma. Rev. Mod. Phys. {\bf 80}, 1083  (2008).
%F. von Oppen, Ya. Peng, F. Pientka. {\it Topological superconducting phases in one dimension}. (Oxford University Press, Oxford, 2015)].

In \cite{Grush-KrylSymmetry2016,Grush-KrylJNPCS2017,Taylor2016,Semiconductor2018}
based on a quasi-relativistic graphene model, two-dimensional (2D) chiral Majorana-like
states in graphene have been described by a high energy $\vec k\cdot \vec p$ Hamiltonian.
Spin-dependent effects stipulate a pseudospin precession under an action of relativistic quantum exchange.
%
%Анализ таких
%майорановских графеновых моделей становится актуальным в связи с
%открытием of an unconventional superconductivity for twisted
%bilayer graphene at a very small angle $\theta_M$ of rotation of
%one monolayer relative to another
%
An analysis of such Majorana-like graphene models becomes relevant in connection
with the discovery of an unconventional superconductivity for twisted
bilayer graphene at a very small angle $ \theta_M $ of rotation of one
monoatomic layer (monolayer) relative to another one \cite{superconductivity2018Graphene}.
A feature of the unconventional superconductivity is accompanying
insulator states such as flat bands being Dirac cone-like  bands
%типа дираковских конусообразных bands
with zero Fermi velocity at  $\theta_M$.  In the case of graphene monolayer without strain a phenomenological
tight-binding model of the graphene superlattice with interlayer
interaction of the graphite type predicts such flat bands at $\theta_M$ only \cite{Bistritzer2011}
 but unfortunately parameters
 %параметры
 of this  non-realistic model can not be adapted to experimental data.
 %нереалистичной модели нельзя адаптировать к экспериментальным данным.

In the paper we   investigate a vortex dynamics of charge carriers in
the quasirelativistic graphene model and its approximations using a high-energy $\vec k\cdot \vec p$
Hamiltonian.
%
%Метод
%вильсоновских незамкнутых петель будет использован, чтобы изучить
%связь между топологией of Brillouin zone, степенью нарушения
%симметрии of band structure, spin-orbital coupling, and different
%типов  resonances in the graphene model.
%
The Wilson non-closed loop method to characterize band-structure topology through
holonomy is used to study the relationship between
the topology of the Brillouin zone, the symmetry breaking
of the band structure, spin-orbital coupling, and different types of resonances
in the graphene model.

\section{Theoretical background}
Graphene is a 2D semimetal hexagonal monolayer,
which is comprised of two trigonal  sublattices $A,\ B$.
Semi-metallicity of graphene  is provided by delocalization
of %Полуметалличность обеспечивается делокализацией
$\pi(\mbox{p}_z)$-electron orbitals  on a  hexagonal crystal cell. % as shown
%на гексагогальной кристаллической ячейке как показано
%in fig.~\ref{color-models}a.
 Since the energies of relativistic terms
 $\pi^* (D_{3/2}) $ and $\pi (P_{3/2}) $ of a hydrogen-like atom are equal each other
 \cite{Fock} there is an indirect exchange % косвенный обмен
 through  %через
 $d$-electron states to break a dimer. Therefore,  a quasirelativistic model monolayer graphene,
besides the configuration with three dimers per the cell,
also has a configuration with two dimers and one broken  conjugate double bond per the cell.
%, as  it is shown
%  кроме конфигурации с
%тремя димерами на ячейку, также holds конфигурацией с двумя димерами и одной разорванной двойной
%сопряженной связью на ячейку, как показано in fig.~\ref{color-models}b.
%Using an approach  \cite{Taylor2016} grounded  on the Dirac--Hartree-Fock self-consistent field approximation
%with the goal to reveal %выявить
%effects of relativistic quantum exchange,  %. Within the framework of the secondary-quantized field approach, one can show that
The high-energy $\vec k\cdot \vec p$ Hamiltonian of a quasiparticle
 in the sublattice, for example, $A$ %of the hexagonal graphene lattice  generalizes the massless Dirac fermion  one and
 reads
\begin{eqnarray}
\left[\vec \sigma \cdot \vec p +  \vec \sigma \cdot \hbar \left(\vec K_B - \vec K_A\right)
\right]\left| \psi^*_{BA}\right>  %\nonumber \\
-{i^2\over c}\Sigma_{AB}\Sigma_{BA}\widehat \psi^\dagger _{-\sigma_A}\left|0,-\sigma \right>
= E_{qu}
\widehat \psi^\dagger_{-\sigma_A} \left|0,-\sigma \right>, %\nonumber \\
\label{Majorana-like-form} %\\
\end{eqnarray}
\begin{eqnarray}
 \left| \psi^*_{BA}\right> = \Sigma_{BA}\widehat \psi^\dagger _{-\sigma_A}\left|0,-\sigma \right>
\ \ \
\label{MwFun}
\end{eqnarray}
where  $\widehat \psi^\dagger _{-\sigma_A}\left|0,-\sigma \right>$
is a spinor  wave function (vector in the Hilbert space), $\vec \sigma =\{\sigma_x, \sigma_y\}$
is the 2D vector of the Pauli matrixes, $\vec p=\{p_x, p_y\}$ is the
2D momentum operator,
 $\Sigma_{AB}$, $\Sigma_{BA}$ are
relativistic exchange operators for sublattices $A, B$
respectively; $i^2\Sigma_{AB}\Sigma_{BA}$ is an unconventional
Majorana-like mass  term for a  quasiparticle in the sublattice $A$,
$\left| \psi^*_{BA}\right>$ is a spinor wave function of
quasiparticle in the sublattice $B$,  $\vec K_A(\vec K_B)$ denote the
graphene Dirac point  (valley) $\vec K$($\vec K'$) in the Brillouin zone; $c$ is the speed of light. A  small term
$\hbar \vec \sigma\cdot\left( \vec K    _B - \vec K_A \right)\sim
{h\over a}$ in eq.~(\ref{Majorana-like-form}) is a spin--valley-current coupling.
One can see that the term with conventional mass in (\ref{Majorana-like-form}) is absent.
Since the exchange operators transform a wave function from sublattice $A$ into $B$ and visa versa
in accord with eq.\eqref{MwFun} the following expression holds
%\begin{widetext}
$$
\left|\tilde\psi^*_{BA}\right\rangle=\Sigma_{BA}\Sigma_{AB}\Sigma_{BA}
\left|\psi_A\right\rangle =\Sigma_{BA}^2\Sigma_{AB} \psi_A+
\Sigma_{BA}  [\Sigma_{AB},\Sigma_{BA}]\left|\psi_A\right\rangle= \Sigma_{BA}^2\{\Sigma_{AB} +
\Sigma_{BA}^{-1}[\Sigma_{AB},\Sigma_{BA}]\} \left|\psi_A \right\rangle.
$$
%\end{widetext}
Since the latter can be written in a form
$\left| \tilde \psi^*_{BA} \right\rangle= \Sigma_{BA}^2  \left|\tilde \psi_{A} \right\rangle
$
%and taking into account of
%$$\tilde {\psi}_{A}= \Sigma'_{AB} {\psi}_{A},  \Sigma'_{AB}= \Sigma_{BA}^{-1}\Sigma_{AB}  \Sigma_{BA}.$$
one gets the following property of the exchange operator matrix:
\begin{equation}
\left|\tilde \psi^*_{BA} \right\rangle
\equiv \alpha ^{-1}  \Sigma_{BA} \left|\tilde\psi_{A} \right\rangle = \Sigma_{BA}^2 \left|\tilde\psi_{A}\right\rangle,
\label{property}
\end{equation}
with some parameter $\alpha$.
%NEW TEXT !!!!!!!!!!!!!!!!!!!!!!!!!
Due to the property \eqref{property}, one can transform the equation
(\ref{Majorana-like-form}) to a following form:
%\begin{widetext}
\begin{equation}
\begin{split}
\left( \Sigma^{-1}_{BA}\vec \sigma \Sigma_{BA}  \right) \cdot \left( \Sigma^{-1}_{BA} \left( \vec p +\hbar (\vec K_B-\vec K_A)
\right) \Sigma_{BA}   \right) \Sigma^{-1}_{BA} \Sigma^2_{BA}
\left(   \Sigma^{-1}_{BA}\hat{\psi}^{\dagger}_{-\sigma_A}
\Sigma_{BA} \right)  \left( \Sigma^{-1}_{BA} \left | 0, -\sigma\right>\right) - i^2{\alpha\over c} \\
\times
\left(\Sigma^{-1}_{BA}\Sigma_{AB}\Sigma_{BA}^2\right)\Sigma_{BA}
\left(\Sigma^{-1}_{BA}\hat{\psi}^{\dagger}_{-\sigma_A}\Sigma_{BA} \right)  \left( \Sigma^{-1}_{BA} \left | 0, -\sigma\right>\right)
%\\
=
\Sigma^{-1}_{BA} E_{qu}\Sigma_{BA}\left( \Sigma^{-1}_{BA}\hat{\psi}^{\dagger}_{-\sigma_A}
\Sigma_{BA} \right)  \left( \Sigma^{-1}_{BA} \left | 0, -\sigma\right>\right). \label{new-gapped-Majorana1}
\end{split}
\end{equation}
%\end{widetext}
Let us introduce the following notations
\begin{equation}
\begin{split}
\vec \sigma^{\prime}_{AB}= \Sigma_{BA}^{-1} \vec \sigma  \Sigma_{BA},\
\vec p\,^{\prime}_{BA}= \Sigma_{BA}^{-1} \,\vec p\,  \Sigma_{BA}, \ \
\vec {K^{\prime}}^{BA}_{B} -\vec {K^\prime}^{BA}_{A} = \Sigma_{BA}^{-1} (\vec K_B  - \vec K_A )\Sigma_{BA},\\
M_{BA}= i^2 \alpha \Sigma_{BA} \Sigma_{AB}, \ M_{AB}= i^2 \alpha
\Sigma_{AB} \Sigma_{BA}, \label{mass-operator} \ \
\hat{ \psi^\prime}^{\dagger}_{-\sigma^\prime_A} = \Sigma^{-1}_{BA}\hat{\psi}^{\dagger}_{-\sigma_A}
\Sigma_{BA} ,\
\left | 0, - \sigma^\prime\right>=\Sigma^{-1}_{BA} \left | 0, -\sigma\right> %\label{transformed-wave-function}
.
\end{split}
\end{equation}
Then  eq.~(\ref{new-gapped-Majorana1}) can be rewritten as
%\begin{widetext}
\begin{equation}
\begin{split}
\left[\vec \sigma^{\prime}_{AB}\cdot \left(\vec p\,^{\prime}_{BA}+\hbar(\vec {K^{\prime}}_B^{BA}-\vec {K^{\prime}}_A^{BA})
\right)  - {1\over c} M^{\prime}_{AB}
\right]\Sigma_{BA}\hat{ \psi^\prime}^{\dagger}_{-\sigma^\prime_A}\left| 0, - \sigma^\prime\right>=
\hat v_F^{-1}E_{qu} \Sigma_{BA}\hat{ \psi^\prime}^{\dagger}_{-\sigma^\prime_A}\left| 0, - \sigma^\prime\right>.
\label{new-gapped-Majorana2}
\end{split}
\end{equation}
%\end{widetext}
Here $\hat v_F$ is the Fermi velocity operator: $\hat v_F=\Sigma_{BA} $,
\begin{equation}
M^{\prime}_{AB}= \Sigma_{BA}^{-1}M_{AB}\Sigma_{BA}. \label{transformed-mass}
\end{equation}
The equation %Уравнение
(\ref{new-gapped-Majorana2}) formally is similar to the massless Dirac fermion equation.
%формально схоже с уравнением Дирака.

Let us prove that the mass operators $M_{AB},\ M_{BA}$ %in the Majorana-like equations of motion
remain invariant under an action of exchange interactions \eqref{transformed-mass},
%{\bf \large Theorem. \\}
%\begin{theorem}
namely, the transformed  mass operator $M^{\prime}_{AB}$ \eqref{transformed-mass} for an electron (hole) in
the Majorana mode represents itself % представляет собой
the  mass operator $M_{BA}$ for a hole (electron) in this mode.
%\end{theorem}
%\Proof

%{\\ \bf \large Proof\\ }
Using the property \eqref{property} we transform
eq.~(\ref{Majorana-like-form}) in an another way:
%\begin{widetext}
\begin{equation}
\begin{split}
\left( \Sigma_{BA}\vec \sigma \Sigma^{-1}_{BA}  \right) \cdot \left( \Sigma_{BA} \left( \vec p +\hbar (\vec K_B-\vec K_A)
\right) \Sigma^{-1}_{BA}   \right)  \Sigma_{BA}
\left(   \Sigma_{BA}\hat{\psi}^{\dagger}_{-\sigma_A}
 \left | 0, -\sigma\right>\right) \\ - i^2{\alpha\over c}
\left(\Sigma_{BA}\Sigma_{AB}\right)\Sigma_{BA} \left(   \Sigma_{BA}
\hat{\psi}^{\dagger}_{-\sigma_A}  \left | 0, -\sigma\right>\right)
 =E_{qu}
 \Sigma_{BA}\hat{\psi}^{\dagger}_{-\sigma_A} \left | 0, -\sigma\right>. \label{new-gapped-Majorana3}
\end{split}
\end{equation}
%\end{widetext}
Then due to the property \eqref{property},
$\Sigma_{BA} \left(\Sigma_{BA}\hat{\psi}^{\dagger}_{-\sigma_A}\right) =
{1\over \alpha }\Sigma_{BA} \hat{\psi}^{\dagger}_{-\sigma_A}$ and by  the following notations
\begin{equation}
\vec \sigma_{AB}= \Sigma_{BA} \vec \sigma  \Sigma^{-1}_{BA},\
\vec p_{BA}= \Sigma_{BA} \,\vec p\,  \Sigma^{-1}_{BA},\
\vec K^{BA}_{B} -\vec K^{BA}_{A} = \Sigma_{BA} (\vec K_B  - \vec K_A )\Sigma^{-1}_{BA},
\end{equation}
 eq.~(\ref{new-gapped-Majorana3}) can be rewritten  as
\begin{eqnarray}
\left[\vec \sigma_{AB}\cdot \left(\vec p_{BA} + \hbar (\vec K^{BA}_B - \vec K^{BA}_A)\right)
-{1\over c} M_{BA}\right]
% \nonumber \\ \times
\Sigma_{BA} \hat{\psi}^{\dagger}_{-\sigma_A}  \left | 0, -\sigma\right>
=\hat v_F^{-1}E_{qu}\Sigma_{BA} \hat{\psi}^{\dagger}_{-\sigma_A}\left|0,-\sigma \right>
. %\nonumber \\
\label{gapped-Majorana1}
\end{eqnarray}
Since the operator $\Sigma_{BA}$ acts on vectors %действует на векторы
$\hat{ \psi^\prime}^{\dagger}_{-\sigma^\prime_A}\left| 0, - \sigma^\prime\right>$ and
$ \hat{\psi}^{\dagger}_{-\sigma_A}\left|0,-\sigma \right>$, belonging to the same Hilbert space,
%принадлежащие одному и тому же гильбертовому пространству,
the operator $M_{BA}$ represents itself
%то the operator $M_{BA}$ представляет собой
a result of the transformation \eqref{transformed-mass}:
\begin{equation}
M_{BA} = M^{\prime}_{AB}. \label{transformed-mass1}
\end{equation}
 Owing to the invariance of the operators  %инвариантности операторов
 $M_{BA},\ M_{AB}$ in respect to  %относительно
 the transformation \eqref{transformed-mass}, their eigenvalues are dynamical masses
%их собственные значения являются значениями масс
of the Dirac fermions %фермионов
which these fermions gain in the Majorana-like superposition
%в
%м%айорано-подобной суперпозиции (майорановской электрон-дырочной паре)
(Majorana electron-hole pair).
%\hfill $\blacksquare$

The equation similar to %Уравнение, аналогичное to
(\ref{gapped-Majorana1}), can be also written for the %для
sublattice $B$. As a result, one gets the %получаем
equations of motion for a Majorana bispinor $(\left|\psi_{AB}\right \rangle, \left|\psi^*_{BA}\right \rangle)^T$
 \cite{Grush-KrylSymmetry2016,myNPCS18-2015}:
\begin{eqnarray}
\left[\vec \sigma_{2D}^{BA}\cdot \vec p_{AB} -c^{-1} %\widetilde  {\Sigma_{AB}\Sigma_{BA}}
M_{AB}
\right]\left|
\psi_{AB}\right\rangle =
i {\partial \over \partial t}\left| \psi^*_{BA}\right \rangle , %\nonumber\\
\label{Majorana-bispinor01}  \\
\left[\vec \sigma_{2D}^{AB}\cdot \vec p\,^*_{BA}
-c^{-1}\left( M_{BA} % \widetilde {\Sigma_{BA}\Sigma_{AB}}
\right)^*\right] \left|\psi_{BA}\right\rangle %\nonumber\\
  = - i {\partial \over \partial t}\left|
\psi^*_{AB}\right \rangle . %\nonumber\\
\quad\label{Majorana-bispinor02}
\end{eqnarray}

%Let us choose the electron configuration consisting of the %Выберем электронную конфигурация из
%Kramers doublets of $\pi(\mbox{p}_z)$-electrons in fig.~\ref{color-models}a as initial conditions for the solution
%a в качестве начальных условий решения
%of the system (\ref{Majorana-bispinor01}, \ref{Majorana-bispinor02}).
%According to %Согласно
%(\ref{Majorana-bispinor01}, \ref{Majorana-bispinor02}) the $\pi(\mbox{p}_z)$-electron configuration
%evolve %эволюционирует
%into Majorana states %майорановские состояния
%that the $K_A$-valley Dirac current splits (fission) into Majorana fermions
%$\left|\psi_{AB}\right\rangle$ and $\left| \psi_{BA}\right \rangle$ ($\gamma_2$ and $\gamma_1$ in fig.\ref{color-models}c)
%by  the exchange $\Sigma_{AB}$ on the sublattice $A$, and the $K_B$-valley Dirac current splits (fission) into conjugated %сопряженные
%Majorana fermions
%$\left|\psi^*_{AB}\right\rangle$ and $\left| \psi^*_{BA}\right \rangle$ ($\overline{\gamma}_1$, $\overline{\gamma}_2$
%in fig.\ref{color-models}c) by  the exchange $\Sigma_{BA}$ on the sublattice $B$.
%The conjugated Majorana fermions "annihilate"\ %"аннигилируют"\
%(fuse) with formation of Dirac fermions %c образованием дираковских фермионов
%(two dimers).
%Since the two-dimers-configurations  and the $K_A$- and $K_B$-valley Dirac currents are equivalent
%эквивалентны
%these fusions and fissions braid the Majorana fermions.
%The feature %Отличительная черта
%of the Majorana braiding is that this braiding is realized in the system which is similar to
%doubled %реализуется в системе, подобной удвоенной
%Kitaev's chain.

\section{Band structure and non-Abelian Zak phase simulations}

The system of eqs.~(\ref{Majorana-bispinor01}, \ref{Majorana-bispinor02})
for the stationary case can be approximated by %может быть аппроксимирована
a Dirac-like equation with a "Majorana-force"\ correction in the following way. %следующим образом.
The operator $\Sigma_{AB}^{-1}$ in (\ref{Majorana-bispinor01}) plays a role of Fermi velocity also:
$\hat v'_F=\Sigma_{AB}$. Then one can assume that there is a following expansion
up to a normalization constant $\left<0\right|\hat v_F \left|0 \right>=\left<0\right|\hat v'_F \left|0 \right>$:
%Taking into account that in accord with %cогласно
%\eqref{gapped-Majorana1} the operator $\Sigma_{BA}$ can be considered as
%a Fermi velocity operator %может быть рассмотрен, как оператор скорости Ферми
%$\hat v_F=\Sigma_{AB}^{-1}$, for example, in \eqref{Majorana-bispinor02}, the following expansion  holds
%up to a normalization constant
%с точностью до нормировочного множителя
%$\left<0\right|\hat v_F \left|0 \right>$:
%\begin{widetext}
\begin{eqnarray}
\left| \psi_{AB}\right\rangle =\frac{\Sigma_{AB}\left| \psi^*_{BA}\right\rangle}{\left<0\right|\hat v'_F \left|0 \right>}=
%\propto
\frac{\Sigma_{AB}\Sigma_{BA}}{
\left<0\right|\hat v'_F \left|0 \right>} \left|
\psi_{AB}\right\rangle %\nonumber \\
=\frac{\left\{ \Sigma_{BA}  + \left[\Sigma_{AB}, \Sigma_{BA}\right]\right\}
\left|\psi_{AB}\right\rangle}{ \left<0\right|\hat v_F \left|0
\right>} %
    \nonumber \\
\approx \left\{ 1  + {\left(\Delta \Sigma + \left[\Sigma_{AB}, \Sigma_{BA}\right]\right)
\over \left<0\right|\hat v_F \left|0 \right>}\right\} \left| \psi_{AB}\right\rangle %\nonumber \\
 \label{permutations1}
\end{eqnarray}
%\end{widetext}
where $\left[\cdot, \cdot \right]$ denotes the
commutator, $\Delta \Sigma = \Sigma_{BA} - \Sigma_{AB} $.
Substituting (\ref{MwFun}, \ref{permutations1}) into the right-hand side of the equation %уравнение
\eqref{Majorana-bispinor02}, one gets the basic Dirac-like equation with a
%small
 "Majorana-force"\ correction of an order of energy difference of quantum exchange for two
graphene sublattices:
\begin{eqnarray}
\left[ \vec \sigma_{2D}^{AB}\cdot \vec p_{BA}-c^{-1} M_{BA} %\widetilde {\Sigma_{AB}\Sigma_{BA}}
\right]
\left| \psi^*_{BA}\right\rangle % \nonumber \\
 = \tilde E\left\{ 1 +{\left(\Delta
\Sigma + \left[\Sigma_{AB}, \Sigma_{BA}\right]\right)\over
\left<0\right|\hat v_F \left|0 \right>} \right\}\left|
\psi^*_{BA}\right \rangle \label{Majorana-bispinor1}
\end{eqnarray}
where $\tilde E=E/\left<0\right|\hat v_F \left|0 \right>$.
The exchange interaction term $\Sigma_{rel}^{x}$  is determined as
\cite{NPCS18-2015GrushevskayaKrylovGaisyonokSerow}
%\begin{widetext}
\begin{eqnarray}
\Sigma_{rel}^{x}\left(
\begin{array}{c}
\widehat {\chi } ^{\dagger}_{_{-\sigma_{_A}} }(\vec r) \\%[0.6mm]
\widehat {\chi }^\dagger _{\sigma_{_B}}(\vec r)
\end{array}
\right)\left|0,-\sigma \right> \left|0,\sigma \right>
=
 \left(
\begin{array}{cc}
0&  \Sigma_{AB}
\\
\Sigma_{BA} & 0
\end{array}
\right)
%\nonumber\\&\times
\left(
\begin{array}{c}
\widehat {\chi }^{\dagger}_{-\sigma_{_A} } (\vec r) \\%[0.6mm]
\widehat {\chi} ^\dagger _{\sigma_{_B}}(\vec r)
\end{array}
\right)\left|0,-\sigma \right> \left|0,\sigma \right>  \label{exchange}
, \\ %[3mm]
\Sigma_{AB}
\widehat {\chi }^\dagger _{\sigma_{_B}}(\vec r)\left|0,\sigma \right>
%\nonumber \\
=
\sum_{i=1}^{N_v\,N}\int { d \vec r_i}
\widehat {\chi }^\dagger _{\sigma_i{^B}}(\vec r)\left|0,\sigma \right> \Delta _{AB}
\langle 0,-\sigma_i|{\widehat \chi}^\dag_{-\sigma_i^A} (\vec r_i)
V(\vec r_i -\vec r)
{\widehat \chi}_{-\sigma_B}(\vec r_i)|0,-\sigma_{i'}\rangle , \ \
\label{Sigma-AB}
\\ %[3mm]
 \Sigma_{BA}
\widehat {\chi }^{\dagger}_{_{-\sigma_{_A}} } (\vec r)
\left|0,-\sigma \right>
%\nonumber\\
=\sum_{i'=1}^{N_v\,N}\int { d \vec r_{i'}}
\widehat {\chi }^{\dagger}_{_{-\sigma_{i'}^A} } (\vec r)
\left|0,-\sigma \right> \Delta _{BA}
\langle 0,\sigma_{i'}|{\widehat \chi}^\dag_{\sigma_{i'}^B} (\vec r_{i'})
V(\vec r_{i'} -\vec r)
{\widehat \chi}_{_{\sigma_A}}(\vec r_{i'})|0,\sigma_i\rangle . \ \
\label{Sigma-BA}
\end{eqnarray}
%\end{widetext}
Here interaction ($2\times 2$)-matrices $\Delta _{AB}$ and $\Delta _{BA}$ are gauge fields
(or components of a gauge field). Vector-potentials for
these gauge fields are determined by the phases % фазами
$\alpha_{ 0}$ and $\alpha_{\pm, k}$, $k=1,\ 2,\ 3$ of $\pi(\mbox{p}_z)$-electron wave functions
$\psi_{\mbox{\small p}_z}(\vec r)$ and $\psi_{\mbox{\small p}_z, \pm \vec \delta_k}(\vec r)$,
$k=1,\ 2,\ 3$ respectively that %within a quasi-relativistic Dirac--Hartree--Fock approach
the exchange interaction
$\Sigma_{rel}^{x}$ (\ref{exchange}) in accounting of the nearest lattice neighbours  for a tight-binding
approximation %for the  system of equations \eqref{Majorana-bispinor01}, \eqref{Majorana-bispinor02}
reads %(\ref{Sigma-AB01}, \ref{Sigma-BA01}) )
\cite{Taylor2016,myNPCS18-2015,NPCS18-2015GrushevskayaKrylovGaisyonokSerow}:
%\begin{widetext}
\begin{eqnarray}
 \Sigma_{AB}%\approx \Sigma_{AB}(\delta_i q_i)
 ={1\over \sqrt{2}(2\pi)^{3}}
e^{-\imath (\theta_{k_{A}}-\theta_{K_B})}%\nonumber\\ \times
\sum_{i=1}^{3} \exp\{\imath [\vec K^i_{A} - \vec q_i ] \cdot \vec
\delta_i\} \int  V(\vec r) d {\vec r}
\nonumber \\
 \times \left(
\begin{array}{cc}
\sqrt{2}\psi_{\mbox{\small p}_z} (\vec r )
\psi^*_{\mbox{\small p}_z, - \vec \delta_i} (\vec r )
 &
\psi_{\mbox{\small p}_z} (\vec r ) [\psi^*_{\mbox{\small
p}_z}(\vec r)
+ \psi^*_{\mbox{\small p}_z, - \vec \delta_i}(\vec r)]\\
\psi^*_{\mbox{\small p}_z, - \vec \delta_i} (\vec r)
[\psi_{\mbox{\small p}_z, \vec \delta_i}(\vec r)+
\psi_{\mbox{\small p}_z}(\vec r)] & {[\psi_{\mbox{\small
p}_z, \vec \delta_i}(\vec r)+ \psi_{\mbox{\small p}_z}(\vec
r)] [\psi^*_{\mbox{\small p}_z}(\vec r)
+\psi^*_{\mbox{\small p}_z, - \vec \delta_i}(\vec r)] \over
\sqrt{2}}
\end{array}
\right)
%\nonumber \\= {1\over \sqrt{2}(2\pi)^{3}} e^{-\imath (\theta_{k_{A}}-\theta_{K_B})}%\nonumber\\ \times
%\sum_{i=1}^{3} \exp\{\imath [\vec K^i_{A} - \vec q_i ] \cdot \vec \delta_i\} \Sigma_{AB}^{nd} \ \ \ \ \ \ \
, \label{Sigma-AB3}
\\ %[0.3cm]
%\end{eqnarray}\begin{eqnarray}
\Sigma_{BA}%\approx   \Sigma_{BA} (\delta_i q_i)
= {1\over \sqrt{2}(2\pi)^{3}} e^{-\imath
(\theta_{K_A}-\theta_{K_B})}
%\nonumber \\ \times
\sum_{i=1}^{3} \exp\{\imath [\vec K^i_{A} - \vec q_i ] \cdot \vec
\delta_i\}  \int  V(\vec r) d {\vec r}
\nonumber \\
 \times \left(
\begin{array}{cc}
{[\psi_{\mbox{\small p}_z, \vec \delta_i}(\vec r)+
\psi_{\mbox{\small p}_z}(\vec r)] [\psi^*_{\mbox{\small
p}_z}(\vec r) +\psi^*_{\mbox{\small p}_z, - \vec
\delta_i}(\vec r)] \over \sqrt{2}}
 &
- \psi^*_{\mbox{\small p}_z, - \vec \delta_i} (\vec r )
[\psi_{\mbox{\small p}_z, \vec \delta_i}(\vec r)+
\psi_{\mbox{\small p}_z}(\vec r)]
\\
- \psi_{\mbox{\small p}_z} (\vec r ) [\psi^*_{\mbox{\small
p}_z}(\vec r) + \psi^*_{\mbox{\small p}_z, - \vec
\delta_i}(\vec r)] & \sqrt{2}\psi_{\mbox{\small p}_z} (\vec
r ) \psi^*_{\mbox{\small p}_z, - \vec \delta_i} (\vec r)
\end{array}
\right)
%\nonumber \\ ={1\over \sqrt{2}(2\pi)^{3}} e^{-\imath (\theta_{K_A}-\theta_{K_B})}
%\nonumber \\ \times
%\sum_{i=1}^{3} \exp\{\imath [\vec K^i_{A} - \vec q_i ] \cdot \vec
%\delta_i\} \Sigma_{BA}^{nd}
 \label{Sigma-BA3}
\end{eqnarray}
%\end{widetext}
%??? не определены переменные --$theta_{K_A}$???
where the origin of the reference frame is located at a given site on the sublattice $A$($B$),
 $V(\vec r)$ is the three-dimensional (3D) Coulomb potential,
%$\psi_{\mbox{\small p}_z} (\vec r)$ is the wave function of p$_z$-electron, and
designations $\psi_{\mbox{\small p}_z,\ \pm \vec \delta_i}(\vec r)$,
$\psi_{\mbox{\small p}_z,\ \pm \vec \delta_i}(\vec r_{2D})\equiv \psi_{\mbox{\small p}_z}(\vec r\pm \vec \delta_i)$,
$i=1,2,3$ refer to %относятся к
atomic orbitals  of p$_z$-electrons with 3D radius-vectors $\vec r\pm \vec \delta_i$ in the neighbor lattice sites
%в соседних узлах
$\vec \delta_i$, nearest to the reference site%?????!!!! ближайших к узлу, размещенному в точке отчета,
%is  introduced in the following way: %Here  введено обозначение
;
 $\vec r\pm \vec \delta_i$ is the p$_z$-electron 3D-radius-vector.
%The wave functions are defined up to  a phase multiplier.
%Let us denote the phases of the wave functions %волновых функций
%$\psi_{\mbox{\small p}_z}(\vec r)$ and $\psi_{\mbox{\small p}_z, \pm \vec \delta_k}(\vec r)$,
%$k=1,\ 2,\ 3$ as $\alpha_{ 0}$ and $\alpha_{\pm, k}$, $k=1,\ 2,\ 3$ respectively.
%A set of these phases is a four-dimensional (4D) phase $\alpha ^\mu _{\pm}$, $\mu=0,\ \ldots,\ 3$, whose components
%компоненты которой
%determine  a vector potential of the gauge field.
Elements of the matrices $\Sigma_{AB}$ and $\Sigma_{BA}$ %in the following way:
%????????????????
%matrix elements
%Матричные элементы
%$\Sigma_{ij,AB(BA)}$
include bilinear %, содержат билинейные по
combinations of the wave functions so that %комбинации волновых функций, так что
their phases  $\alpha_{ 0}$ and $\alpha_{\pm, k}$, $k=1,\ 2,\ 3$
enter into $\Delta_{AB}$ and $\Delta_{BA}$ from % входят в
(\ref{Sigma-AB} and \ref{Sigma-BA}) in the form
%в виде
%
%??? !!! написать правильное поведение при калибровочном преобразовании
%
\begin{eqnarray}
%\left(\psi + \delta\psi\right) \left(\psi +
%\delta\psi\right)^*_{\pm \vec \delta_k} %\nonumber \\
%=
\left|\psi_{\mbox{\small p}_z}\right| \left|\psi_{\mbox{\small p}_z,\ \pm \vec \delta_k}\right|
 \exp\left\{\imath \left( \alpha_0 - \alpha_{\pm,
k}\right)\right\} % \nonumber \\
\equiv \left|\psi_{\mbox{\small p}_z}\right| \left|\psi_{\mbox{\small p}_z,\ \pm \vec \delta_k}\right|
\Delta _{\pm,k}
 %\exp\left\{\imath \Delta \alpha _{\pm, k} \right\}
. \label{phase_variation}
\end{eqnarray}
%
%\begin{eqnarray}
% \left<\hat S(\alpha_\mu)\psi_{\pm \vec
%\delta_k}\right|
%\hat O
%\left| \hat S(\alpha_\mu)\psi\right>=
%
%\left<\psi|\right|\hat O \left|\psi _{\pm \vec
%\delta_k}\right> \exp\left\{\imath \Delta \alpha _{\pm, k} \right\}
%\end{eqnarray}
%
Therefore, an effective number %Эффективное число
$N$ of flavors in our gauge field theory is equal to 3. %В силу трансляционной симметрии и
%гексагональной симметрии кристаллической ячейки эффективно имеются две независимые фазы для каждой кристаллической ячейки
%Because of the translational symmetry and the hexagonal symmetry
%of the crystal cell, there are effectively two independent phases $c_{\pm}$ for each crystal cell (see Appendix A1).
Then owing to %В силу
translational symmetry %and choice %выбору
%of  the wave function with accuracy up to the %с точностью до
%phase multiplier,
we  determine the  gauge fields $\Delta _{\pm,i}$ in eq.~(\ref{phase_variation})
%through introducing phase multiplier $\exp\left\{\imath \Delta \alpha _{\pm, i} \right\}$
%to the wave function at site $\pm\delta_i$, $i=1,2,3$
in the following form:
\begin{equation}
\label{c-alpha}
\Delta _{\pm,i}(q) = \exp\left(\pm \imath c_\pm (q)({\vec q}\cdot{\vec \delta_i})\right).
\end{equation}
Substituting the relative phases \eqref{c-alpha} %относительные фазы
of particles %electrons
and holes %of exciton %электронов и дырок в экситонe
into \eqref{Sigma-AB3} one gets the exchange interaction operator $\Sigma_{AB}$
\begin{eqnarray}
 \Sigma_{AB} ={1\over \sqrt{2}(2\pi)^{3}}
e^{-\imath (\theta_{k_{A}}-\theta_{K_B})}%\nonumber\\ \times
 \left(
\begin{array}{cc}
\Sigma_{11} &\Sigma_{12}\\
\Sigma_{21}& \Sigma_{22}
\end{array}
\right)  \label{Sigma-AB3-second-approximation}
\end{eqnarray}
with following matrix elements:
%\begin{widetext}
\begin{eqnarray}%
\Sigma_{11}= \sqrt{2} \left\{\sum_j I_{11}^j %e^{-\imath c_-(\vec q)(\vec q  \cdot \vec \delta_j)}
\Delta _{-,j}(q)
\exp\{\imath [\vec K^j_{A} - \vec q ] \cdot \vec \delta_j\} \right\},
%\label{Sigma-AB11-second-approximation} \\
%
\Sigma_{12}=   \left\{\sum_j \left( I_{12}^j +  I_{11}^j
% e^{-\imath c_- (\vec q)(\vec q  \cdot \vec \delta_j)}
\Delta _{-,j}(q)
 \right)
\exp\{\imath [\vec K^j_{A} - \vec q ] \cdot \vec \delta_j\}
\right\} , %\nonumber \\
\label{Sigma-AB12-second-approximation}\\
\Sigma_{21}=   \left\{\sum_j \left(I_{21}^j %e^{\imath (c_+(\vec q) - c_-(\vec q))(\vec q  \cdot \vec \delta_j)}
\Delta _{+,j}(q)\Delta _{-,j}(q)
+  I_{11}^j %e^{-\imath   c_-(\vec q)(\vec q \cdot \vec \delta_j)}
\Delta _{-,j}(q)
\right) \exp\{\imath [\vec K^j_{A} - \vec q]
\cdot \vec \delta_j\}
\right\}, \quad \ \ \ \label{Sigma-AB21-second-approximation}\\
\Sigma_{22}= {1 \over \sqrt{2}}   \left\{\sum_j \left( I_{22}^j %e^{\imath c_+(\vec q) (\vec q  \cdot \vec \delta_j)}
\Delta _{+,j}(q)
 +
I_{12}^j+ I_{21}^j % e^{\imath (c_+(\vec q) - c_-(\vec q))(\vec q  \cdot \vec \delta_j)}
\Delta _{+,j}(q)\Delta _{-,j}(q)
%\right.\right.\nonumber \\ \left. \left.
+I_{11} ^j %e^{- \imath c_-(\vec q))(\vec q \cdot \vec \delta_j)}
\Delta _{-,j}(q)
\right) \exp\{\imath [\vec K^j_{A} - \vec q
] \cdot \vec \delta_j \}
\right\}\quad \ \ \ \label{Sigma-AB22-second-approximation}%\\
\end{eqnarray}
where $
I_{n_i m_k}^j = \int  V(\vec r)  \psi_{\mbox{\small p}_z+n_i\vec \delta_j}
{\psi^*}_{\mbox{\small p}_z-m_k\vec \delta_j}  \ d\vec r
$, $i,k=1,2$; $(n_1,m_1)=(0,1)$, % for $I_{11}$,
$(n_1,m_2)=(0,0)$, % for $I_{12}$,
$(n_2,m_1)=(n_2,m_2)=(1,1)$. %, %for $I_{21}$,
%$(n_2,m_2)=(1,1)$. % for $I_{22}$.
%\begin{eqnarray} I_{11} = \int  V(\vec r)  \psi^{(0)}_{\mbox{\small p}_z}{\psi^*}^{(0)}_{\mbox{\small p}_z-\vec \delta_j}
% \ d\vec r , \ \
% I_{12} = \int  V(\vec r)  \psi^{(0)}_{\mbox{\small p}_z}{\psi^*}^{(0)}_{\mbox{\small p}_z} \ d\vec r ,
%\label{Sigma-AB3-second-approximation_Int12}\\
% I_{21} = \int  V(\vec r)  \psi_{\mbox{\small p}_z+\vec \delta_j}{\psi^*}_{\mbox{\small p}_z-\vec \delta_j}  \ d\vec r ,\ \
% I_{22} = \int  V(\vec r)  \psi_{\mbox{\small p}_z+\vec \delta_j}{\psi^*}_{\mbox{\small p}_z} \ d\vec r \ .
%\label{Sigma-AB3-second-approximation_Int12}
%\end{eqnarray}
%\end{widetext}
There are  similar formulas for $\Sigma_{BA}$.
%
%where $ \psi^{(0)}_{\mbox{\small p}_z} $ is the p$_z$-electron orbital.
%We rewrite it in the matrix form as:
%Аналогичный расчет дает
%$ \Sigma_{BA}(\delta_i q_i)$ (\ref{Sigma-BA3})
% in the second order approximation:
%\begin{eqnarray}
% \Sigma_{BA} ={1\over \sqrt{2}(2\pi)^{3}}
%e^{-\imath (\theta_{k_{A}}-\theta_{K_B})}%\nonumber\\ \times
% \left(
%\begin{array}{cc}
%\Sigma_{11}^{BA} &\Sigma^{BA}_{12}\\
%\Sigma_{21}^{AB}& \Sigma^{BA}_{22}
%\end{array}
%\right),  \label{Sigma-BA3-second-approximation}\\%
%\Sigma_{11}^{BA}=\Sigma_{22}^{AB},\
% \label{Sigma-AB11-second-approximation}\\
%\Sigma_{12}^{BA} =-\Sigma_{21}^{AB}  , \ % \label{Sigma-AB12-second-approximation}\\
%\Sigma_{21}^{BA} =-\Sigma_{12}^{AB} , \ % \label{Sigma-AB21-second-approximation}\\
%\Sigma_{22}^{BA} = \Sigma_{11}^{AB}.
%\label{Sigma-BAall-second-approximation}
%\ \ \label{Sigma-BA22-second-approximation}
%\\
%\end{eqnarray}

Now, neglecting %пренебрегая
the  mass term, we can find the solution of %искать решение
the equation \eqref{Majorana-bispinor1} by the successive approximation technique %method of successive approximations
% методом последовательных приближений
as:
%\begin{widetext}
\begin{eqnarray}
\vec \sigma_{2D}^{BA}(\Delta_{\pm, i})\cdot \vec p_{AB}  (\Delta_{\pm, i})\left|
\psi_{AB}\right\rangle +{E^{(0)} \left(\Delta \Sigma (\Delta_{\pm, i})+ \left[\Sigma_{AB} (\Delta_{\pm, i}),
\Sigma_{BA}(\Delta_{\pm, i})\right]\right)\over
\left<0|\hat v_F|0\right>^2 }\left| \psi_{AB}\right \rangle =
{E^{(1)} \over \hat v_F}\left| \psi_{AB}\right \rangle . %\nonumber \\
\label{variational-Majorana-bispinor}
\end{eqnarray}
%\end{widetext}
%Here $\Delta \Sigma,\  \Sigma_{AB}, \Sigma_{BA}$ are determined by the expressions
%(\ref{Sigma-AB3-second-approximation}--\ref{Sigma-AB3-second-approximation_Int12}).
%
%As the exchange operators $\sigma_{2D}^{BA}$ enter
Accordingly to (\ref{c-alpha}) eigenvalues $E^{(1)}_i, \ i=1,\ 2$ of \eqref{variational-Majorana-bispinor} and,
accordingly, eigenvalues $E_i, \ i=1,\ldots, 4$ of  the $4\times 4$ Hamiltonian
(\ref{Majorana-bispinor01}, \ref{Majorana-bispinor02})   are functionals of $c_\pm%_\alpha
$.
%Для устранения произвола в выборе фазовых множителей $c_\pm$  in \eqref{c-alpha} необходима калибровка полей фаз.
To eliminate arbitrariness in the choice of phase factors
$ c_ \pm $ %in \eqref{c-alpha},
one needs a gauge condition for the gauge fields. % must be calibrated.
%\subsection{Second-order approximation}
The eigenvalues $E_i, \ i=1,\ldots, 4$ are real because the system of equations
(\ref{Majorana-bispinor01}, \ref{Majorana-bispinor02})   is
transformed to  Klein--Gordon--Fock equation \cite{Grush-KrylSymmetry2016}.
Therefore we %to solve a eigenproblem for the Hamiltonian (\ref{Majorana-bispinor01}, \ref{Majorana-bispinor02}) will
impose the gauge condition %Калибровочное условие
 as a  requirement on the absence of imaginary parts in the eigenvalues
$E_i, \ i=1,\ldots, 4$ of  the Hamiltonian
(\ref{Majorana-bispinor01}, \ref{Majorana-bispinor02}):
%of \eqref{variational-Majorana-bispinor}. This condition can be written as a system of two equations of the form
\begin{equation}\label{sys}
\Im m(E_{i})=0, \ i=1,\ldots , 4.
\end{equation}
%Direct solution of this system turns out to be unstable for some
%specific points in the momentum space. Instead, for every point
To satisfy the condition (\ref{sys}) in the momentum space we  minimize  a  function $f(c_+%_\alpha
, c_-%_\alpha
)=\sum_{i=1}^4\left|\Im m \ E_{i}\right|$   absolute minimum of which coincides with the solution
of the system
(\ref{sys}). For the mass case band structures for the sublattice Hamiltonians are the same. Therefore neglecting
the mass term the cost function $f={1\over 2} \sum_{i=1}^2\left|\Im m \ E_{i}\right|$. For the non-zero mass case,
we assume the same form of the function $f$ due to smallness of the mass correction.

%\section{Non-abelian Zak phase simulations}

\begin{figure}[tb]
%\begin{center}
\hspace{0.3cm}(a)\hspace{7cm} (b)\\
\hspace*{0.4cm}(c)\hspace{-0.1cm}
\includegraphics[width=3.0cm]{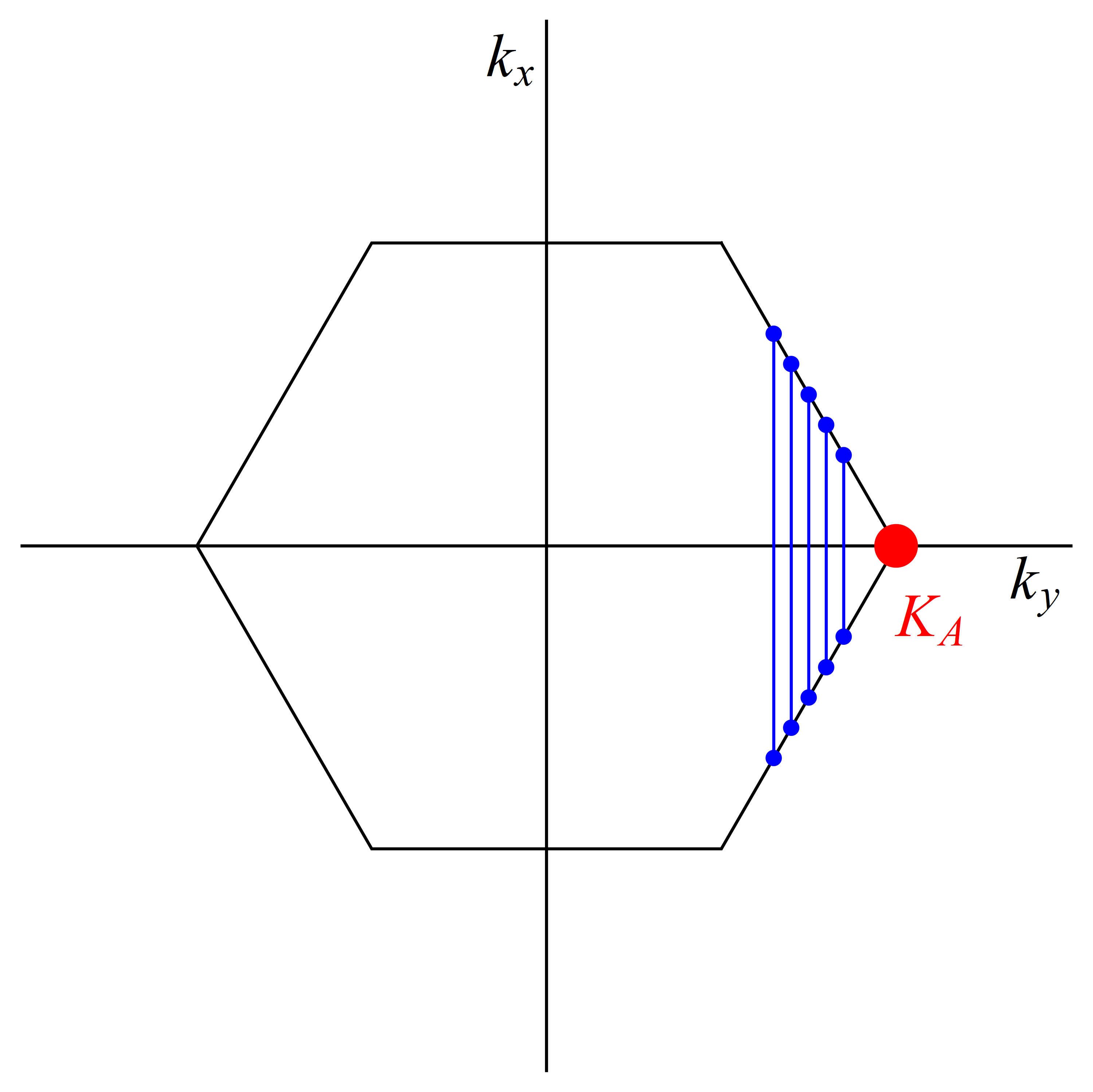} \hspace{2cm} %%% {fig-1_bw.eps for black and white)
%\caption{Scheme of loops for Zak phase simulations in wavevector $\vec k$-space}
%\end{center}
%\end{figure}
%\begin{figure}[htbp]\begin{center}
\includegraphics[width=4.0cm]{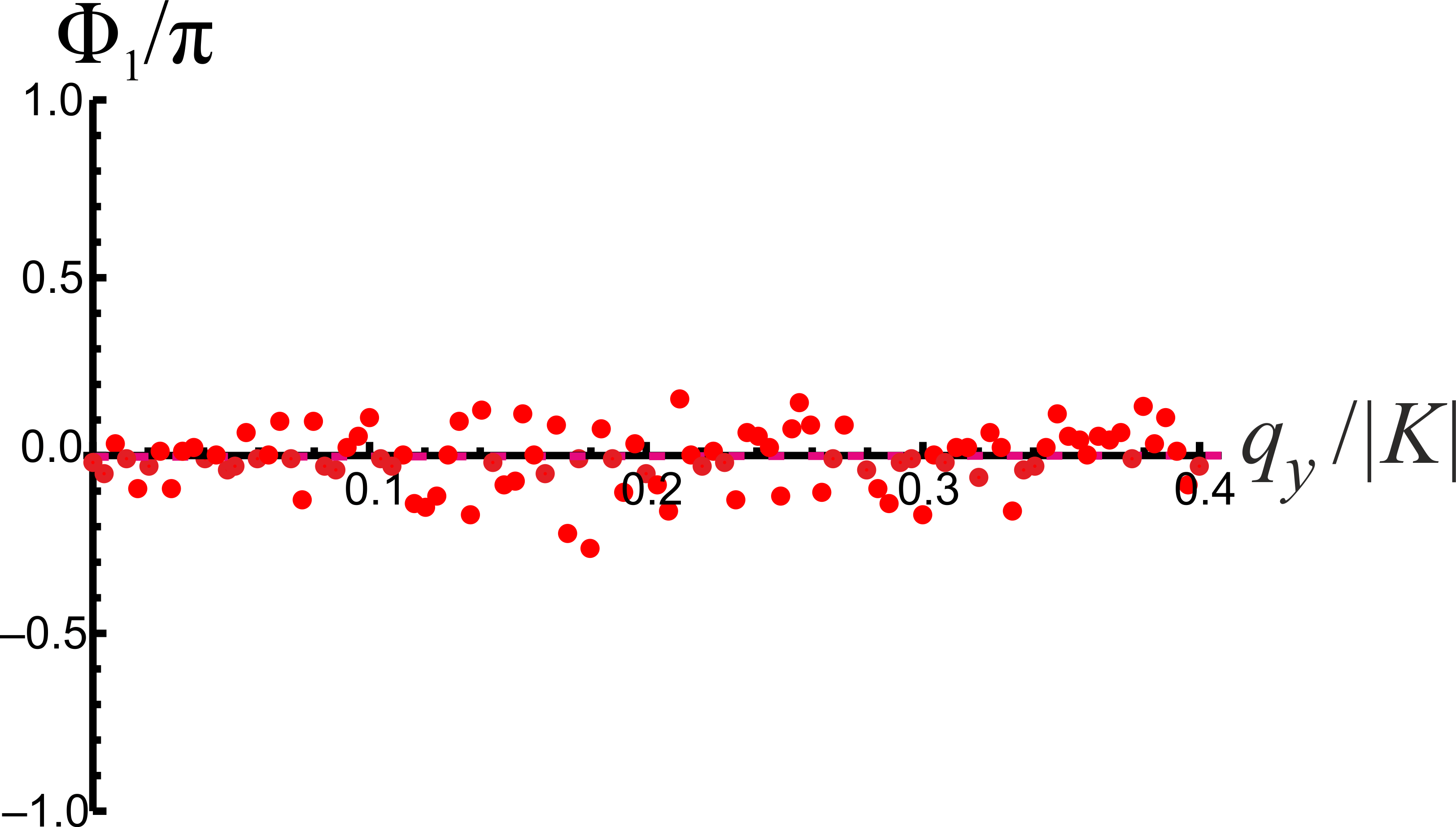} %%% {fig-1_bw.eps for black and white)
\includegraphics[width=4.0cm]{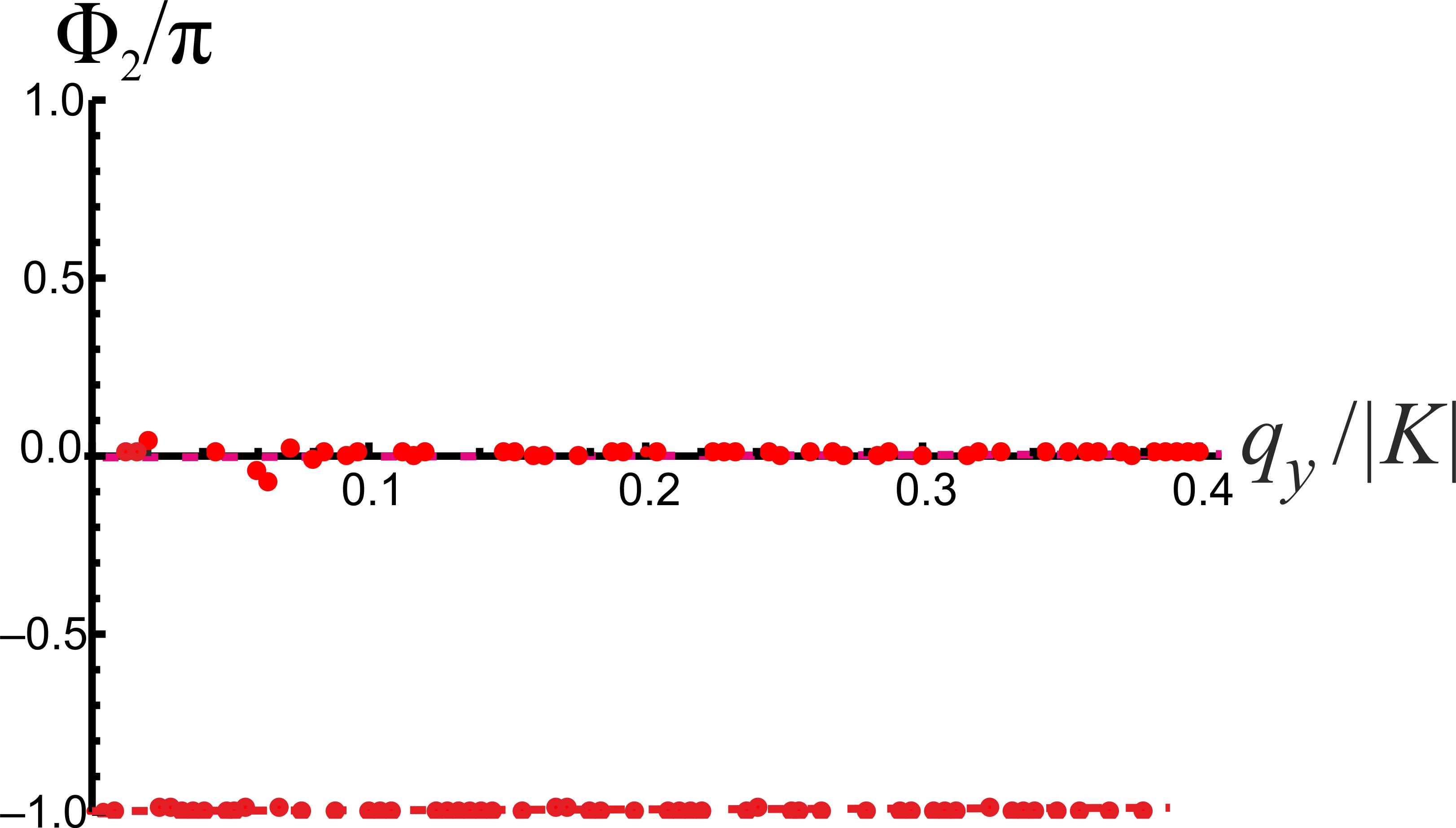} \\%%% {fig-1_bw.eps for black and white)
%(c)\\
\includegraphics[width=1.0cm]{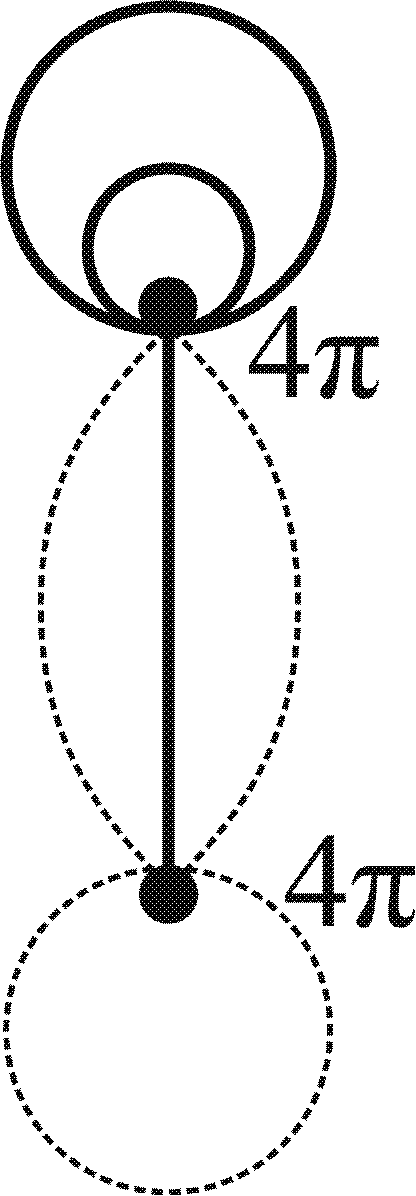}\hspace{4.7cm}
\includegraphics[width=4.0cm]{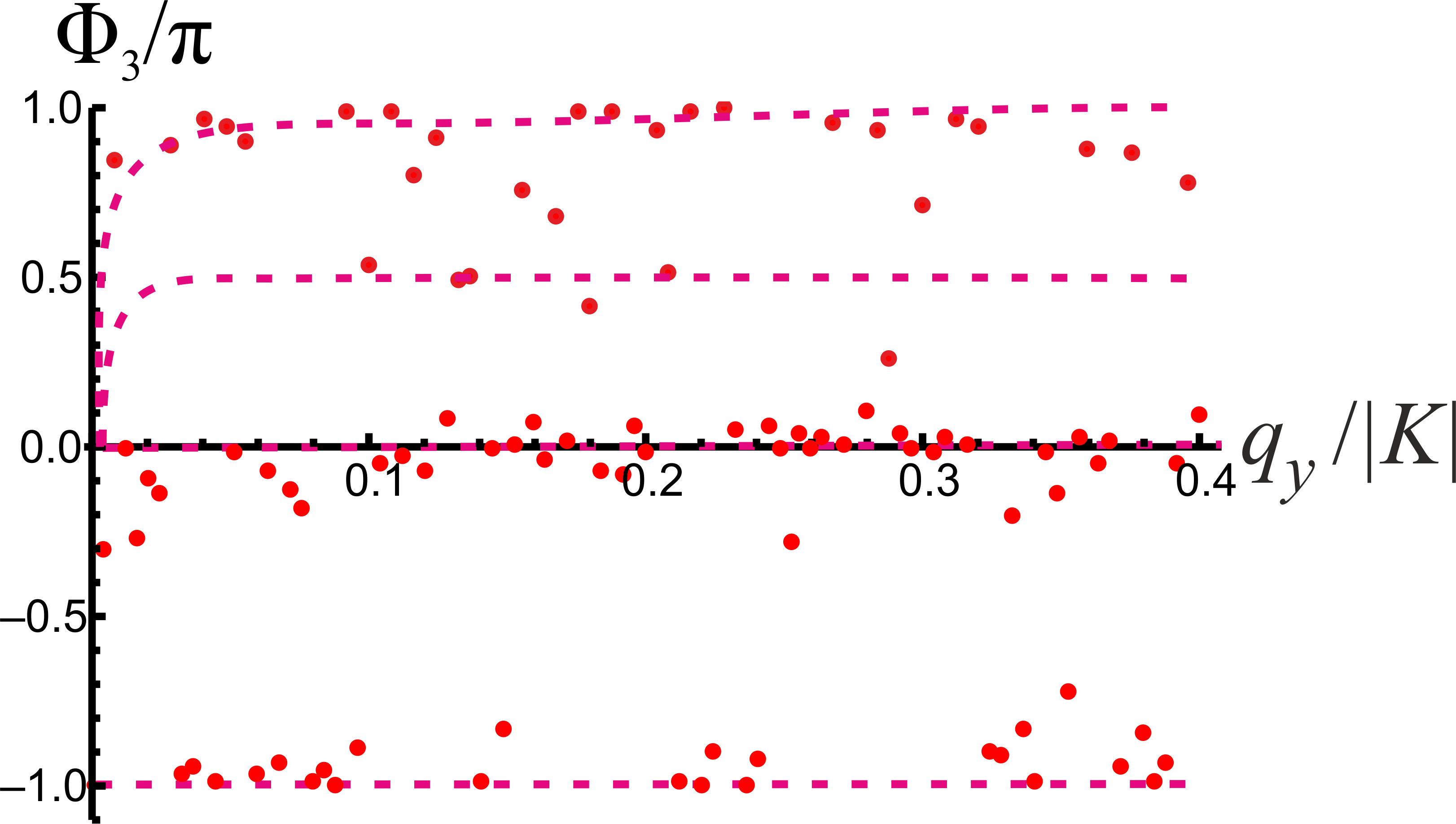}  %%% {fig-1_bw.eps for black and white)
\includegraphics[width=4.0cm]{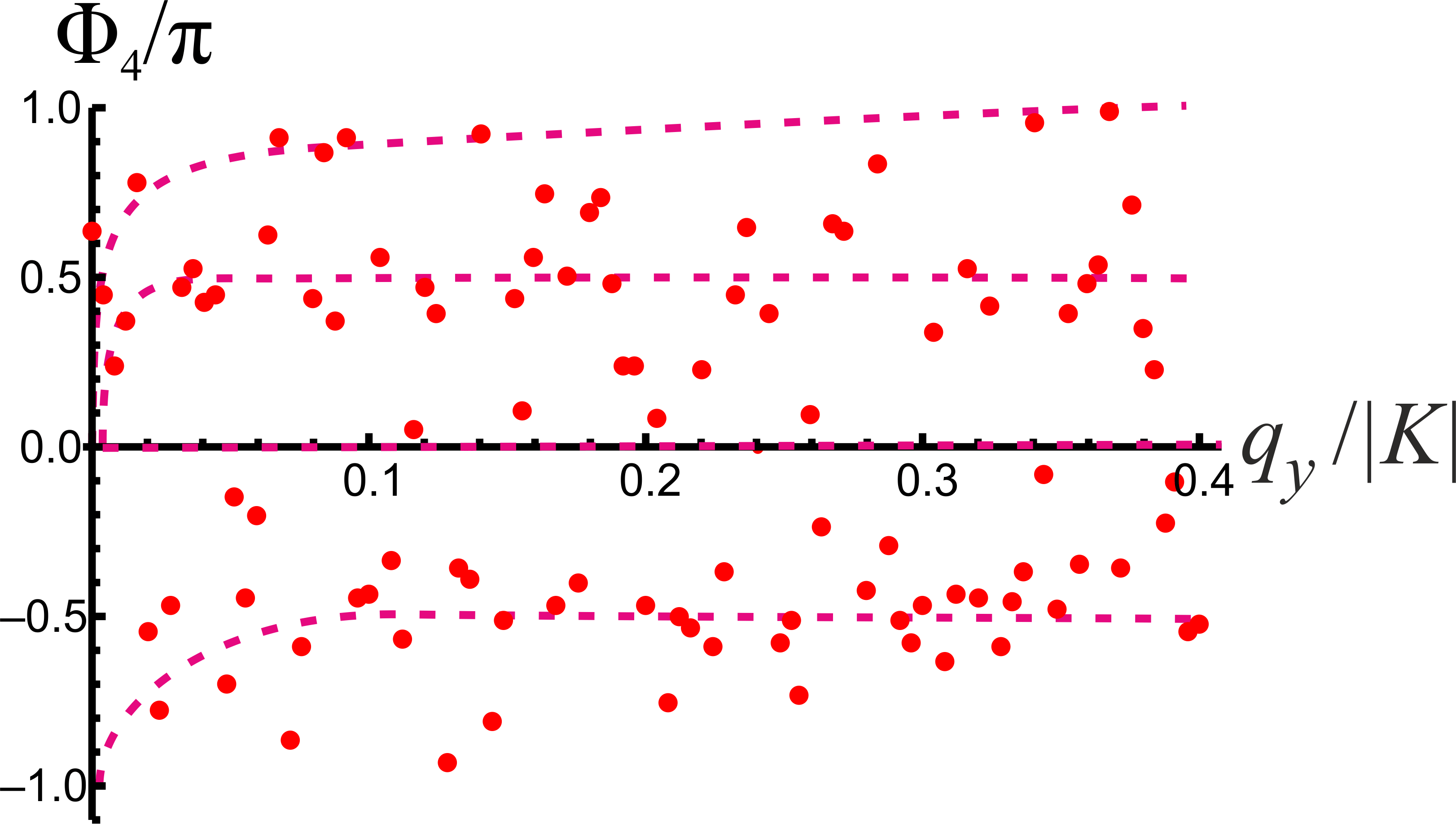} %%% {fig-1_bw.eps for black and white)
\caption{(a) Scheme of Wilson  loops for Zak phase calculation.
(b) Non-Abelian phases $\Phi_1,\ldots,\Phi_4$ of the Wilson-loop eigenvalues in the units of
$\pi$  and (c) sketch of topological defects for
quasi-relativistic model of graphene in approximation of
zero-values of gauge fields, bypass over each contour
% обход по
%каждому контуру
in figure (c) gives phase shift  value
%дает набег фазы
$4\pi$; $\vec q=\vec k -\vec K$.}
%\end{center}
\label{fig1}
\end{figure}

Topological defect pushes out a charge carrier from its location.
The operator of this non-zero displacement
%Топологический дефект
%выталкивает носитель заряда из занимаемой им позиции. Оператор
%этого ненулевого смещения
present a projected position operator ${\cal P}  {\vec r} {\cal P}$
with the projection operator ${\cal P}= \sum_{n=1}^N  \left|\psi_{n,\vec k}\right\rangle \left\langle \psi_{n,\vec k}\right|$
for the occupied subspace of states $\psi_{n,\vec k}(\vec r)$. Here $N$ is a
number of occupied bands, $\vec k$ is a momentum. Eigenvalues of ${\cal P}  {\vec r} {\cal P}$ are called %  называется
Zak phase %фазой
\cite{Zak89}.  The  Zak phase coincides with a phase
\begin{eqnarray}
\gamma_{mn}=i\int\limits_{C(\vec{k})}\left<\psi_{m,\vec{k}}\left|\nabla_{\vec{k}}\right|\psi_{n,\vec{k}}
\right>\cdot d\vec{k}, \quad n,m=1,\ldots, N \label{non-A-ZP}
\end{eqnarray}
of a Wilson loop ${\cal W}^{mn} =\mbox{T }\exp(i  \gamma_{mn}) $
being a path-ordered (T)  exponential with the integral over a closed contour $C(\vec{k})$ \cite{Alexandradinata-et-al2014}.
Let us discretize the Wilson loop  by Wilson lines ${\cal W}_{k_{i+1},k_i}$:
\begin{eqnarray}
{\cal W}=\prod_{i=0}^{N_{\cal W}\to \infty}{\cal W}_{\vec k_{i+1},\vec k_i}=\prod_{i=0}^{N_{\cal W}\to \infty}
\exp\left(-\int_{\vec k_i}^{\vec k_{i+1}}\left<\psi_{m,\vec{q}}\left|{\partial \over \partial \vec q}
\right|\psi_{n,\vec{q}}\right>\cdot d \vec q \right).
 \label{WilsonLoop}
\end{eqnarray}
Here momenta $\vec k_i$, $i=0,1,\ldots, N_W $ form a sequence of the points on a curve
%следуя друг за другом, находятся на кривой
(ordered path), connecting initial and final points in the Brillouin zone:
$\vec k_i=\vec k_0+ \sum_{j=1}^{i}\Delta \vec k_{j,j-1}$, $\Delta \vec k_{j,j-1}=\vec k_{j}-\vec k_{j-1}\to 0$ and
$\vec k_{N_{\cal W}}=\vec k_0$;
$\psi_{n,{\vec k_i}}$, $n=1,\ldots,N$ are eigenstates of a model Hamiltonian.
% соединяющей начало и конец бриллюэновской зоны в заданной направлении.
We perform  the integration by parts  and then
expand the matrix element ${\cal W}^{mn}_{\vec k_{i+1},\vec k_i}$, $\vec k_{i+1}-\vec k_i \to 0$ of the Wilson line with
Bloch waves $\psi_{n,\vec q}(\vec r)$ for our model hamiltonian into series in terms of $\Delta \vec k_{j,j-1}$:
\begin{eqnarray}
&{\cal W}^{mn}_{\vec k_{i+1},\vec k_i}=
\exp\left(-\sum_{\mu}\int d\vec r\int_{k_{\mu,i}}^{k_{\mu,i+1}}
\psi^*_{m,q_\mu}(\vec r){\partial \over \partial q_\mu} \psi_{n,q_\mu}(\vec r) d q_{\mu}\right)=
e^{-\int d\vec r \psi^*_{m,\vec{q}}(\vec r) \psi_{n,\vec{q}}(\vec r)}
\exp\left(\sum_{\mu}\int d\vec r \psi_{n,\overline{q}_\mu}(\vec r)\right. \nonumber \\
&\times \left.
\int_{k_{\mu,i}}^{k_{\mu,i+1}}
{\partial \over \partial q_\mu} \psi^*_{m,{q}_\mu}(\vec r)  d q_{\mu}\right) =
e^{-\delta (q)\delta_{mn}}
\left(1+ \sum_{\mu}\int d\vec r
(\psi^*_{m,k_{\mu,i+1}}(\vec r) - \psi^*_{m,k_{\mu,i}}(\vec r)) \psi_{n,k_{\mu,i}}(\vec r)\right)
,
 \label{Wilson-line}
\end{eqnarray}
where $\delta (q)$ is a Dirac $\delta$-function, $\delta_{mn}$ is the Kronecker symbol,
$\vec k_i \le \vec {\overline q}\le \vec k_{i+1}$. Taking into account that
$e^{-\delta(q)\delta_{mn}}= 1-\delta_{mn}$ and
$1\equiv \sum_{l,l'}\left\langle \psi^*_{l,\vec k_{i}}\left|\right. \psi_{l',\vec k_{i}}\right\rangle$ in the expression
(\ref{Wilson-line}) one gets

\begin{eqnarray}
{\cal W}^{mn}_{\vec k_{i+1},\vec k_i}= (1-\delta_{mn})
\left\langle \psi^*_{m,\vec k_{i}}\left| \right. \psi_{n,\vec
k_{i}}\right\rangle + \left\langle \psi^*_{m,\vec k_{i+1}}\left|
\right. \psi_{n,\vec k_{i}}\right\rangle
 -
\left\langle \psi^*_{m,\vec k_{i}}\left| \right. \psi_{n,\vec k_{i}}\right\rangle
= \left\langle \psi^*_{m,\vec k_{i+1}}\left| \right. \psi_{n,\vec k_{i}}\right\rangle
.
 \label{Wilson-line1}
\end{eqnarray}
In our calculation  of (\ref{WilsonLoop},\ref{Wilson-line1}) a number $N$ of bands is equal to four ($N=4$):
 two electron and hole valent bands and two electron and hole conduction bands.
Instead the closed contour we take a curve $C(\vec{k})$ being one
side of the equilateral triangle of variable size (defined by the
value of $k_y$ component of the wavevector $\vec k$) with the
coordinate system origin in the Dirac $K(K')$-point. The
non-closed loops are shown by vertical lines in fig.~\ref{fig1}a. This
corresponds to the the situation when two vertically placed
endpoints (shown as thick blue points) are equivalent ones  in the Brillouin zone from
the symmetry viewpoint.
The $N$ phases are defined then as arguments of the eigenvalues of
the  Wilson loop.
One chooses $N_{\cal W}$ ($N_{\cal W}=500$) that a "noise"\ in output data  is sufficient small to observe
discrete values of Zak phases.
{\small
\begin{table}[h]
\vspace{-20pt}
\caption{Topological characterization of the graphene models: 1 -- the massless pseudo-Dirac fermion model,
2 and 3 -- the quasirelativistic graphene model in the approximations of zero- and nonzero-gauge field
respectively. Second column from the left: arguments
of the Wilson-loop eigenvalues  ${\cal W}(q_y)$}
\begin{center}
\begin{tabular}{|c|c|}
\hline \hline
                Type of the graphene model       & $\mbox{Arg} \ {\cal W}(q_y)$  \\
\hline \hline
model 1                &$\{0, \ \pm \pi\}$ ;    \\
\hline
model 2               & $\{0, \ \pm \pi/2, \ \pm \pi\}$ ;      \\
\hline
              & $\{0, -\pi/6, -2\pi/6,-3\pi/6,\ldots, -\pi\}$  at $q_y\to 0$,     \\
model 3             & $\{0,  -\pi/4, - \pi/2, -3\pi/4, -\pi \}$ at $q_y>0.2|K|$, \\
                       & $\{0,  -\pi/4, - \pi/2, -3\pi/4, -\pi \}$  and
                       $\{0,  \pi/2,  \pi\}$ at $q_y>0.24|K|$ \\
\hline
\end{tabular}
\end{center}\label{table1}
%\vspace{-12pt}
\end{table}
}

\section{Results and discussion}
\begin{figure}[tb]
%\begin{center}
%\hspace{1cm}
\hspace*{0.5cm}{(a)} \hspace{7.5cm}(c)\\
\hspace*{-8.cm}\includegraphics[width=4.0cm]{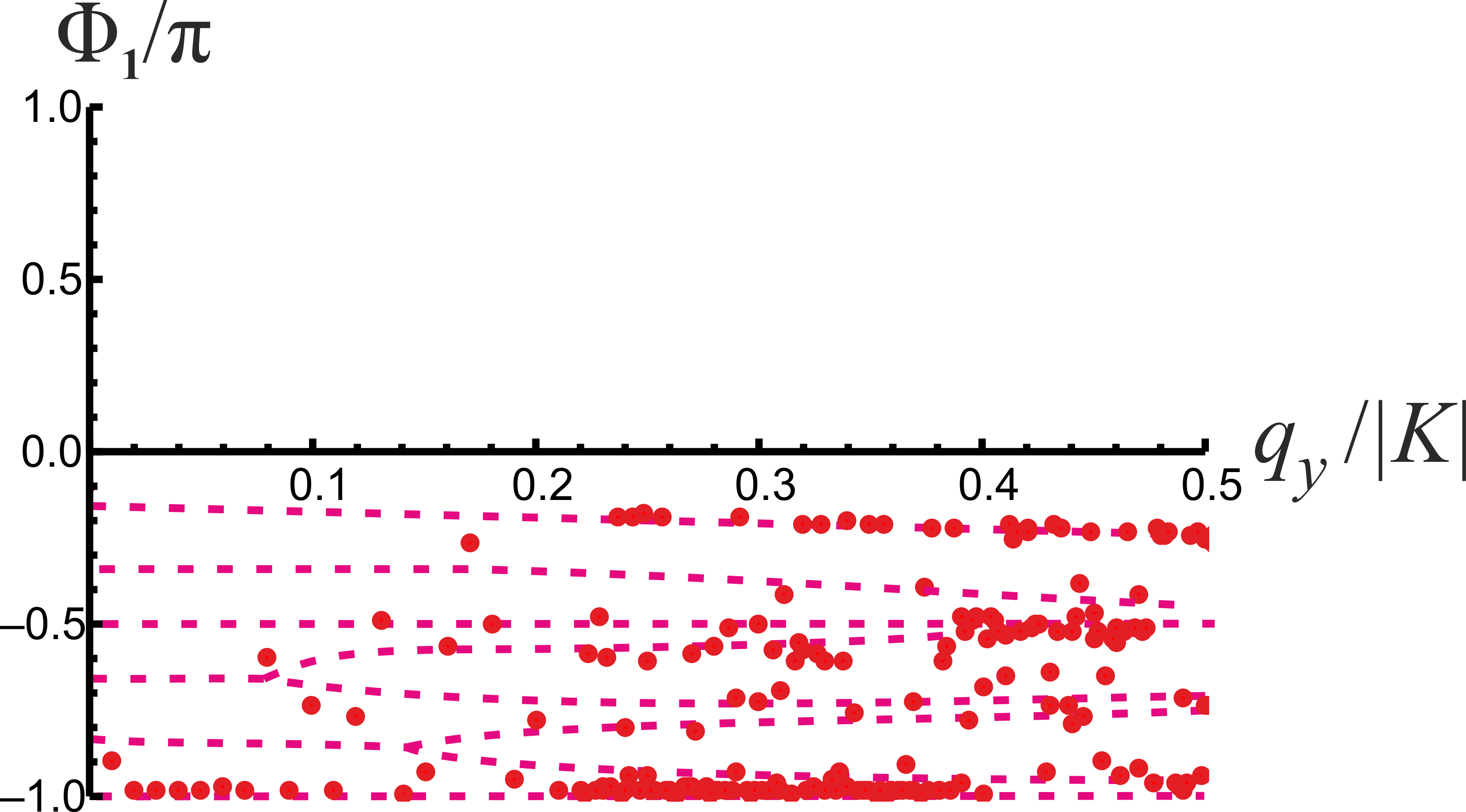} %%% {fig-1_bw.eps for black and white)
%\hspace*{1.cm}
\includegraphics[width=4.0cm]{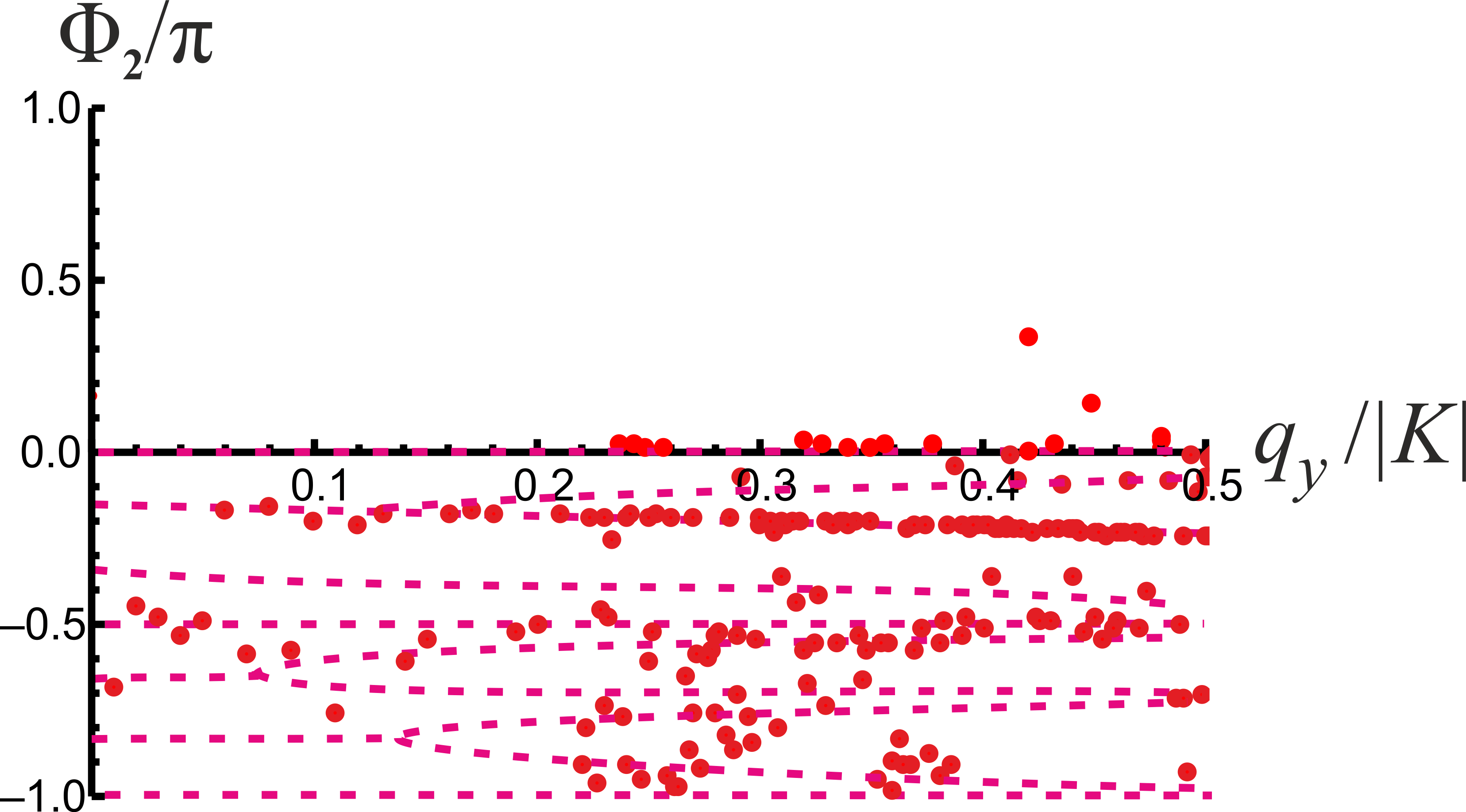}%\includegraphics[width=2.0cm]{topologyZ8.png}
\vspace*{0.5cm}
\\
\hspace*{-8.cm} \includegraphics[width=4.0cm]{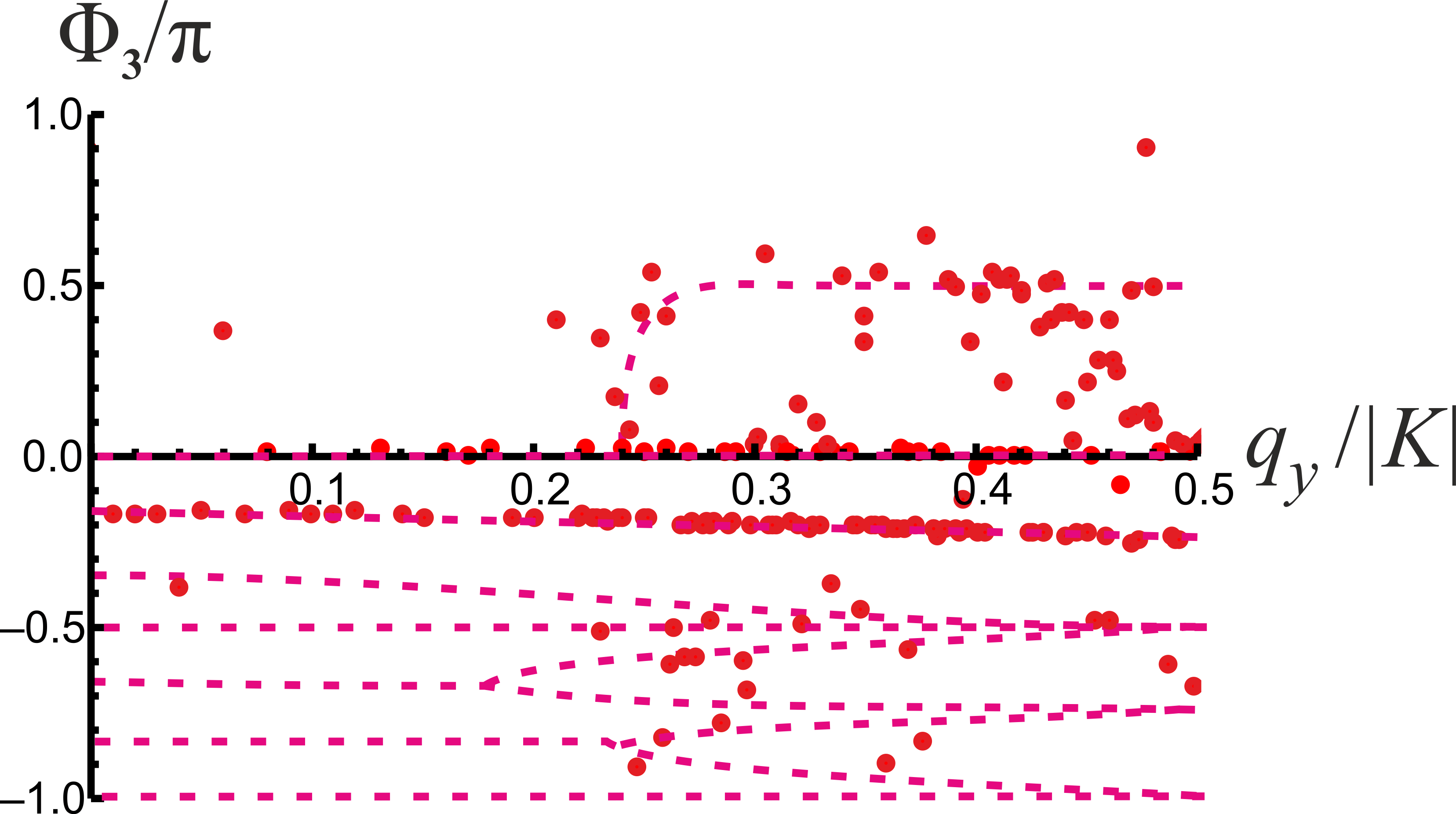} %%% {fig-1_bw.eps for black and white)
\includegraphics[width=4.0cm]{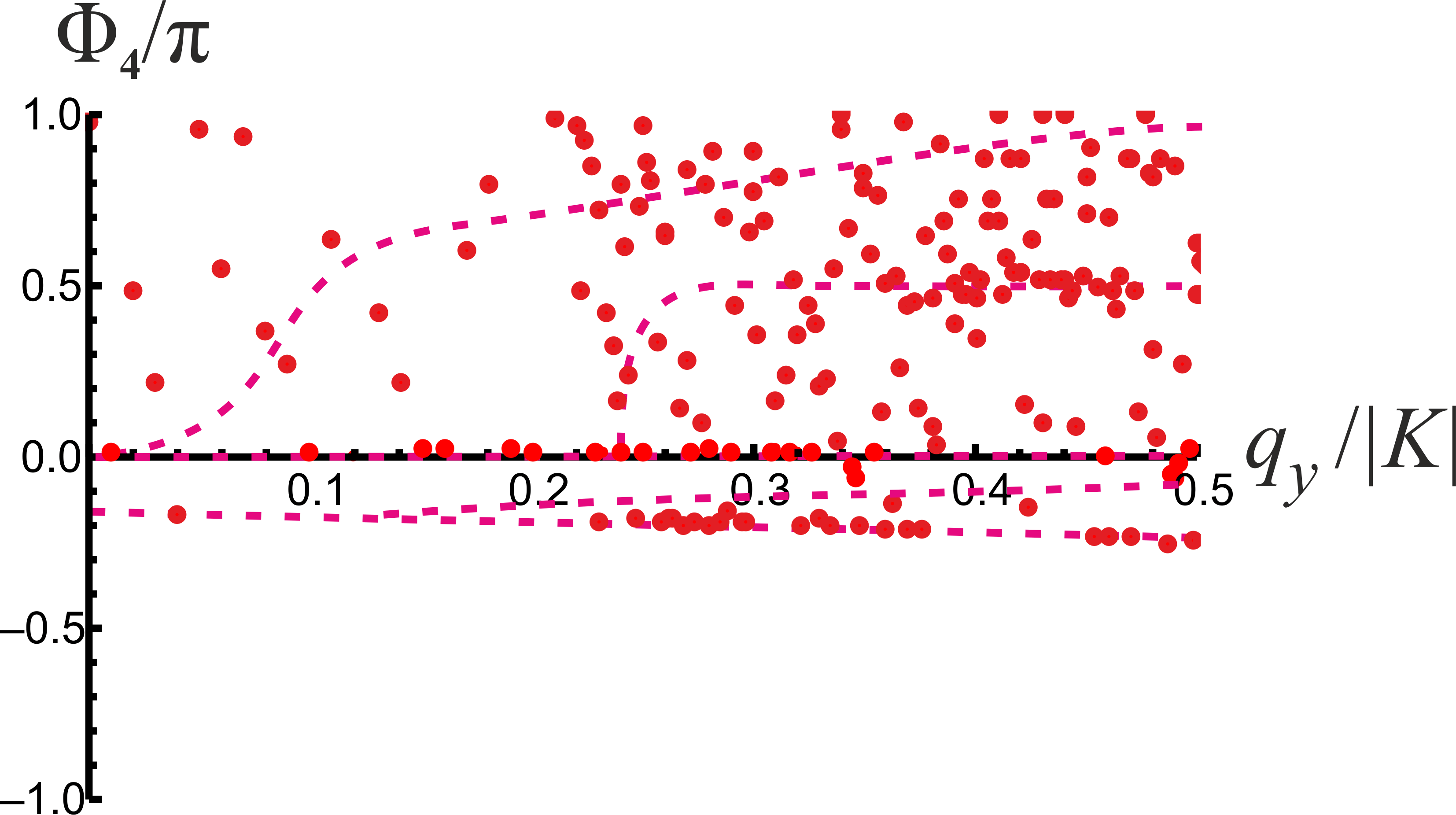} %\hspace*{1cm}(c)
\vspace*{-4.9cm}
\\
%\hspace{1cm} \hspace{8cm}  \vspace*{-2.2cm} \\
%\hspace*{1cm} (b)\includegraphics[width=4.0cm]{topologyZ8.png}
%
%\vspace*{-4.2cm}
 \hspace*{9.4cm} \includegraphics[width=8.3cm]{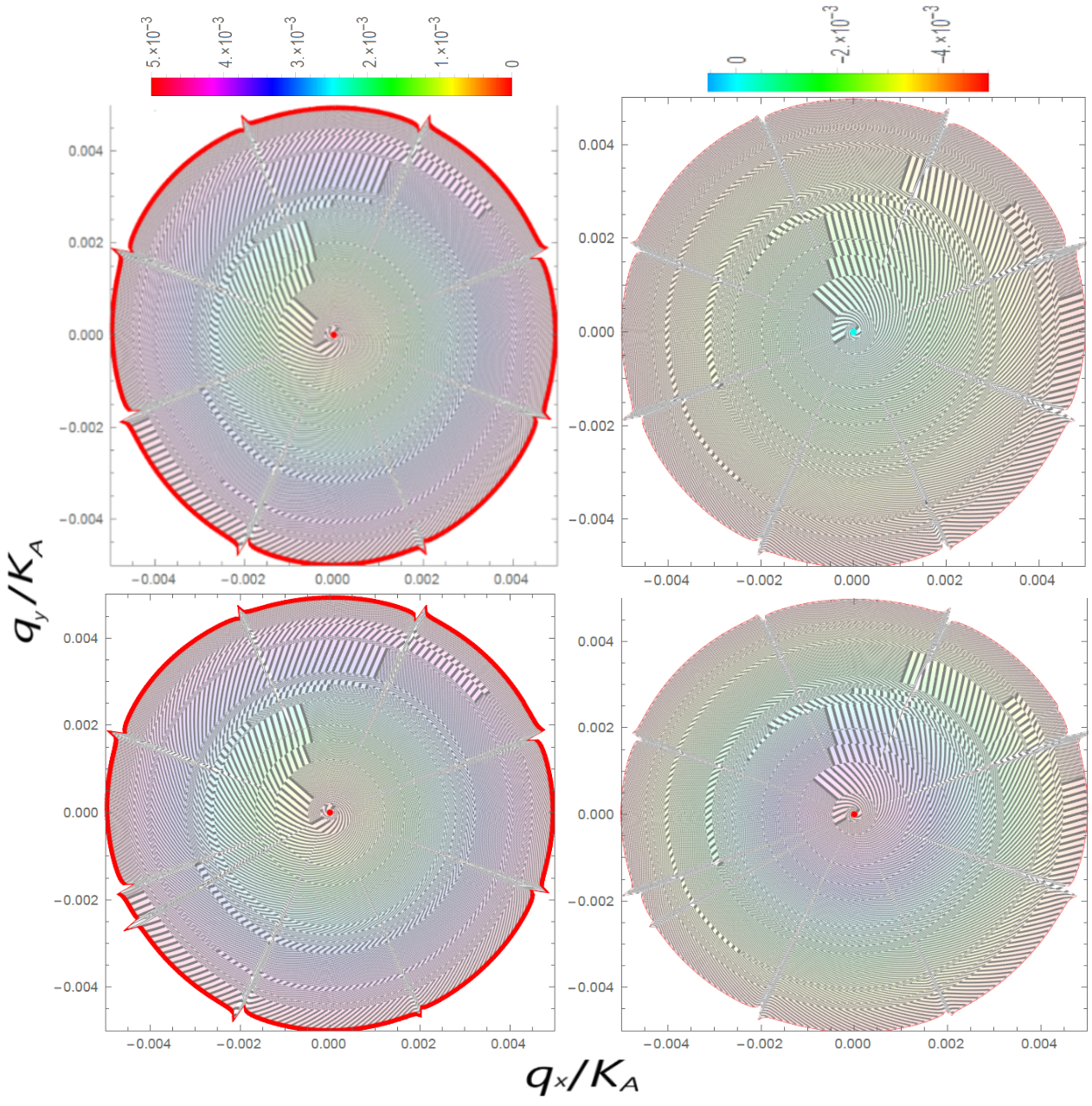}\vspace*{-2.2cm}
 \\
\hspace*{-5.2cm} (b)\\
\hspace*{-3.cm} \hspace*{-5.2cm} \includegraphics[width=2.0cm]{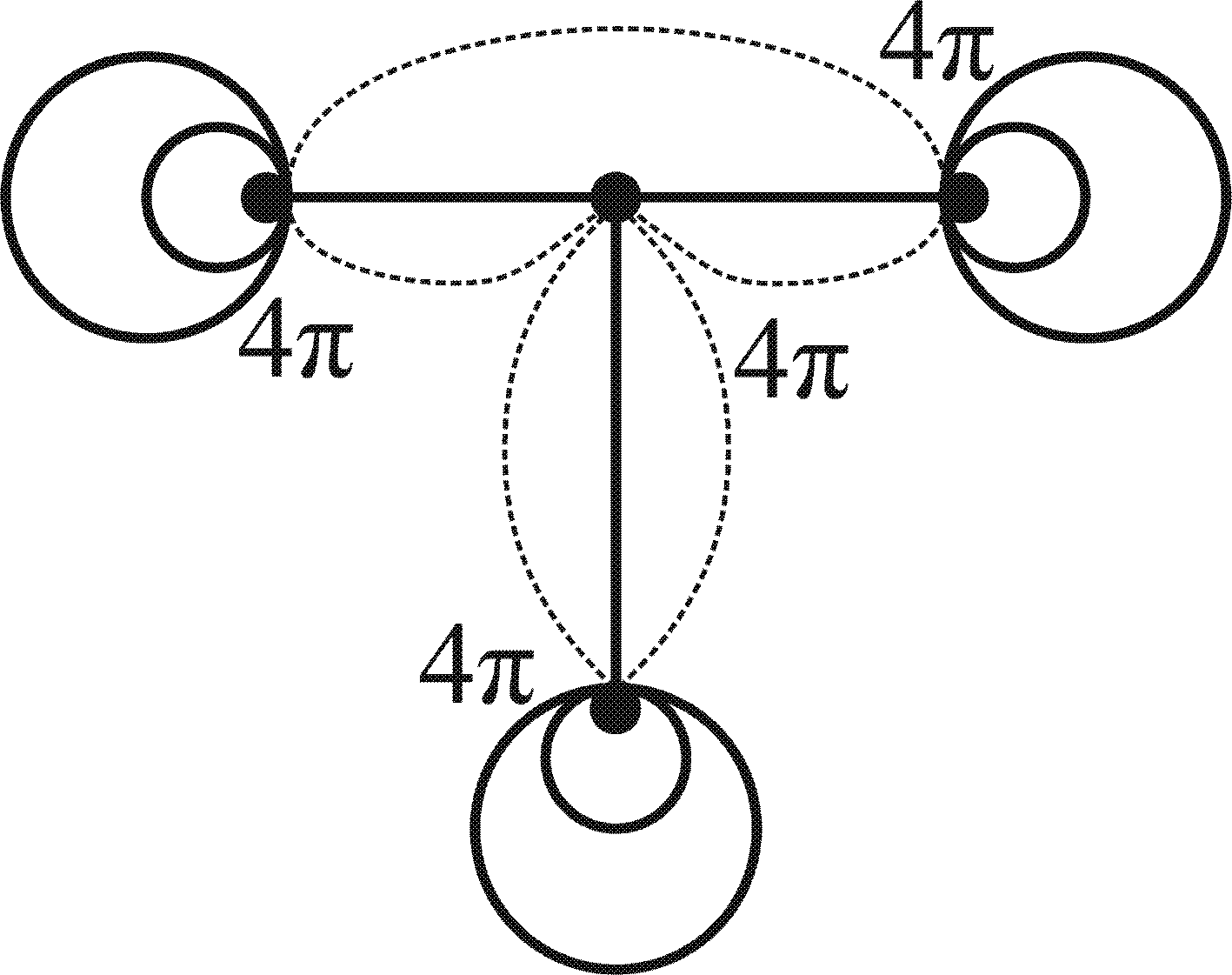} %
\vspace*{1.2cm}\\
%%% {fig-1_bw.eps for black and white)
\caption{(a) Non-Abelian phases $\Phi_1,\ldots,\Phi_4$ of the Wilson-loop eigenvalues in the
units of $\pi$  and (b) sketch of topological defects for
quasi-relativistic model of graphene with non-zero gauge field,
bypass over each contour %обход по  каждому контуру
in figure (b) gives phase shift  value
% дает набег фазы
$4\pi$; $\vec q=\vec k -\vec K$. (c) A vortex texture in contour
plots of electron (left) and hole (right) bands at the non-zero
gauge field,
 bands have been calculated
 %рассчитывались
 on momentum scales %(масштаб):
$q/|K|\sim 0.002$ for non-interacting
%для невзаимодействующих
(up) and interacting
%и взаимодействующих
(down) electron and hole constituted
%  электрона и дырки,
%составляющих
a Majorana particle.
 } \label{fig2}
%\end{center}
\end{figure}

Simulation of the Zak phases has been performed for three
gra\-phe\-ne models, the first one is a massless pseudo-Dirac
fermion model \cite{JPhysB40-2007Semenoff}, the second one is our quasi-relativistic graphene model %  \cite{myNPCS18-2015}
in
an approximation of zero gauge field and the third model is the
same one but with accounting of non-zero gauge field.
  The results are presented in Table~\ref{table1} and figs.~\ref{fig1}b,~\ref{fig2}a.

 % As one can see,
Up to finite accuracy of the numeric method we %indeed
get a discrete set of obtained values of phases for considered models. For the
massless Dirac model, arguments of the Wilson-loop eigenvalues are
equal to $0, \pm \pi$ and are multipliers of $\pi/3$. Hence, due
to hexagonal symmetry of the lattice the first model is
topologically trivial one.

 For the second model, there exist  %существуют два
 two different sets of  the Wilson-loop-arguments eigenvalues, namely,  one set %один набор
 $0, \pm \pi $ in the vicinity  of the Dirac point  %в окрестности дираковской точки
 $K(K')$, the second one $0, \pm \pi/2, \pm \pi $  at  sufficiently high values
 %достаточно больших значениях
 of wavevectors $q_y=k_y - K_{A,y}$  (see fig.~\ref{fig1}b).  In the vicinity  %В окрестности
 of $K(K')$  the arguments of the Wilson-loop eigenvalues  form the same  %образуют такую же
 cyclic group $\mathbb{Z}_2$,  as for the case of  %как в
 the massless pseudo-Dirac model.  This testifies on topological non-triviality
 % Это свидетельствует о топологической нетривиальности
 of the second model. Additional  values  $ \pm \pi/2$ of  the Wilson-loop arguments
and, respectively, two cyclic groups  $\mathbb{Z}_4$ with generators % c образующими
$\pi/2$ or $-\pi/2$ are at values
of %находятся  при значениях
$q_y/|K_A|$ higher than % выше
$0.05$.
% The observed деформация of the cyclic group $\mathbb{Z}_2$ to $\mathbb{Z}_4$
% является следствием возрастания величины спин-орбитального взаимодействия at the large $q_y$.
%
The observed deformation of the cyclic group
$ \mathbb {Z}_2 $ to $ \mathbb {Z}_4 $ is a consequence of the increase
in spin-orbit interaction at the high $ q_y $.
The strong spin-orbital coupling lifts %removes
the degeneration on pseudospin.
%снимает  вырождение по псевдоспину.
Meanwhile Weyl node and antinode emerge.
%возникают.
 Due to the fact that  %Так как
 the cyclic group  is  $\mathbb{Z}_4$, %то
 the Weyl node (antinode) should be a double defects
 %должна быть  удвоенным дефектом
 (in the form of two singular points).
 % (в виде двух особых точек).
 Only in this case,  as it is shown  %Только тогда, как показывает
 in fig.~\ref{fig1}c,  %обход of the удвоенной node (antinode) по контуру c поворотом два раза на угол
 %$4\pi$ дает набег фазы волновой функции на $2\pi$.
 %
 bypassing the doubled node (antinode) along a contour with a double rotation on an angle
 $ 4 \pi $ gives the phase shift for the wave function by $ 2 \pi $.
 %
%Так как the Weyl node (antinode), как любое квантовое фермионное
%состояние, представляют собой Kramers doublet, то найденное его
%удвоение - это результат снятия of degeneration по спину owing to
%the spin-orbital-coupling breaking of time-reversal symmetry.
%
Since the Weyl node (antinode), like any quantum fermion state, is a Kramers doublet,
 then its doubling is a result of splitting spin degeneration
 owing to the spin-orbital-coupling breaking of time-reversal symmetry but without breaking of electron-hole symmetry.
  Resulting  homotopy group  $\mathbb{Z}_2 \times \mathbb{Z}_4$ reveals
a broken symmetry  %нарушенную симметрию
 for the approximation of the quasi-relativistic model with zero gauge field.
Hereafter we show
%Далее мы покажем
that the broken-symmetry group $\mathbb{Z}_2 \times \mathbb{Z}_4$ recovers to
$\mathbb{Z}_8$ of the quasi-relativistic model in the approximation of  with non-zero gauge field.

As simulation results for the third model demonstrate in fig.~\ref{fig2}a,
pathes with the topological Zak phase $- \pi /6$  constitute %составляют
the cyclic groups $\mathbb{Z}_{12}$ in the vicinity %в окрестности
of $K(K')$ at small momenta $q_y$ and additional Majorana Wilson-loop-arguments eigenvalues, multiple to $-\pi/4$ are appeared
at high $q_y$. These results testify that the quasirelativistic
graphene model is topologically nontrivial one in all energy range.
%во всей области энергий.
%Поворот на $- \pi/6$ эквивалентен повороту на $-\pi/2$ due to
%hexagonal symmetry and, correspondingly, электронные и дырочные
%конфигурации в импульсном пространстве ортогональны друг другу.
%
A $( - \pi /6) $ rotation is equivalent to a $ (- \pi/ 2) $ rotation due to hexagonal symmetry and, correspondingly,
the electron and hole configurations in the momentum space are orthogonal to each other.
%Схематическое изображение этой электронно-дырочной конфигурации в виде of a T-shaped trijunction of four peculiar points
%(topological defects% in a form of vortices
%)
%представлено in fig.~???.
%
A schematic representation of this electron-hole configuration in the form of a
T-shaped trijunction of four peculiar points (topological defects)
is presented in fig.~\ref{fig2}b. %An atomic chain %Цепочка атомов с двумя  топологическими дефектами на концах implements a Majorana
%particle.
%
An atomic chain with two topological defects at the ends implements a Majorana particle.
%Therefore the trijunction образуется 3-мя майорановскими particles (modes).
%
Therefore, the trijunction is formed by three Majorana particles (modes) with flavor, a number of flavors $N=3$.
%
%Полный момент импульса $\vec J$  такого
%майораноподобного excitation is equal to $\vec J= \sim_{i=1}^3\vec
%j_i$ с модулем $J=3/2$.
%
The total angular momentum $ \vec J $ of such a Majorana-like excitation
is equal to $ \vec J = \sum_{i = 1}^3 \vec j_i $ with the absolute value $J = 3/2$. Here $\vec j_i$ is angular momentum
%момент  количества движения
of $i$-th Majorana particle, $i=1,2,3$. Majorana and antiMajorana states (excitations) %состояния с
with  $J=3/2$  differ by the projections %различаются проекциями
$J_y=-3/2,-1/2,1/2,3/2$ of the total angular momentum %полного момента
$\vec J$. Therefore the Majorana and antiMajorana excitations confined in the Dirac point
by hexagonal symmetry are fourth-fold degenerated. %четырежды вырождены.
The cyclic group $Z_{12}$ existing at small momenta  $q$, $q\to 0$  testifies that
%hexagonal symmetry удерживает майорановскую mode  в окрестности
%дираковской точки owing to a small spin-orbital coupling.
hexagonal symmetry confines the Majorana mode in the vicinity of the Dirac point owing to  small spin-orbital coupling.
%
%Simulation  обнаруживает восемь право- и лево-закрученных вихрей
%на поверхностях of electron and hole bands that one can observes
%in fig.~???.
%
Simulations discovers eight right- and left-handed vortices on
the surfaces of electron and hole bands that one can observes in fig.~\ref{fig2}c.
%Сопоставим каждой компоненте of the Majorana (anti-Majorana) excitation с значением проекции $J_x$ одно из этих
%вихревых состояний  зонной структуры.
%
We associate each component of the Majorana (anti-Majorana) excitation
with the projection  $ J_y $ one of these vortex states of the band structure.
%Then  совпадение 4-х лево- (право-) закрученных
%вихрей  для the четырежды вырожденных quasirelativistic graphene model bands at small $q$, $q\to 0$ implies вырожденность
%этих вихрей.
Then, the coincidence of the four left-(right-) handed vortices for the fourfold
degenerate quasirelativistic graphene model bands at small $q$, $q\to 0$ implies the degeneracy of these vortices.
%Эти вихревые состояния назовем subreplicas.
We call these vortex states subreplicas.

\begin{figure}[tb]
%\begin{center}
\hspace{3cm}{(a)}\hspace{5cm}{(c)}\\
\hspace*{1.cm}\includegraphics[width=7.0cm]{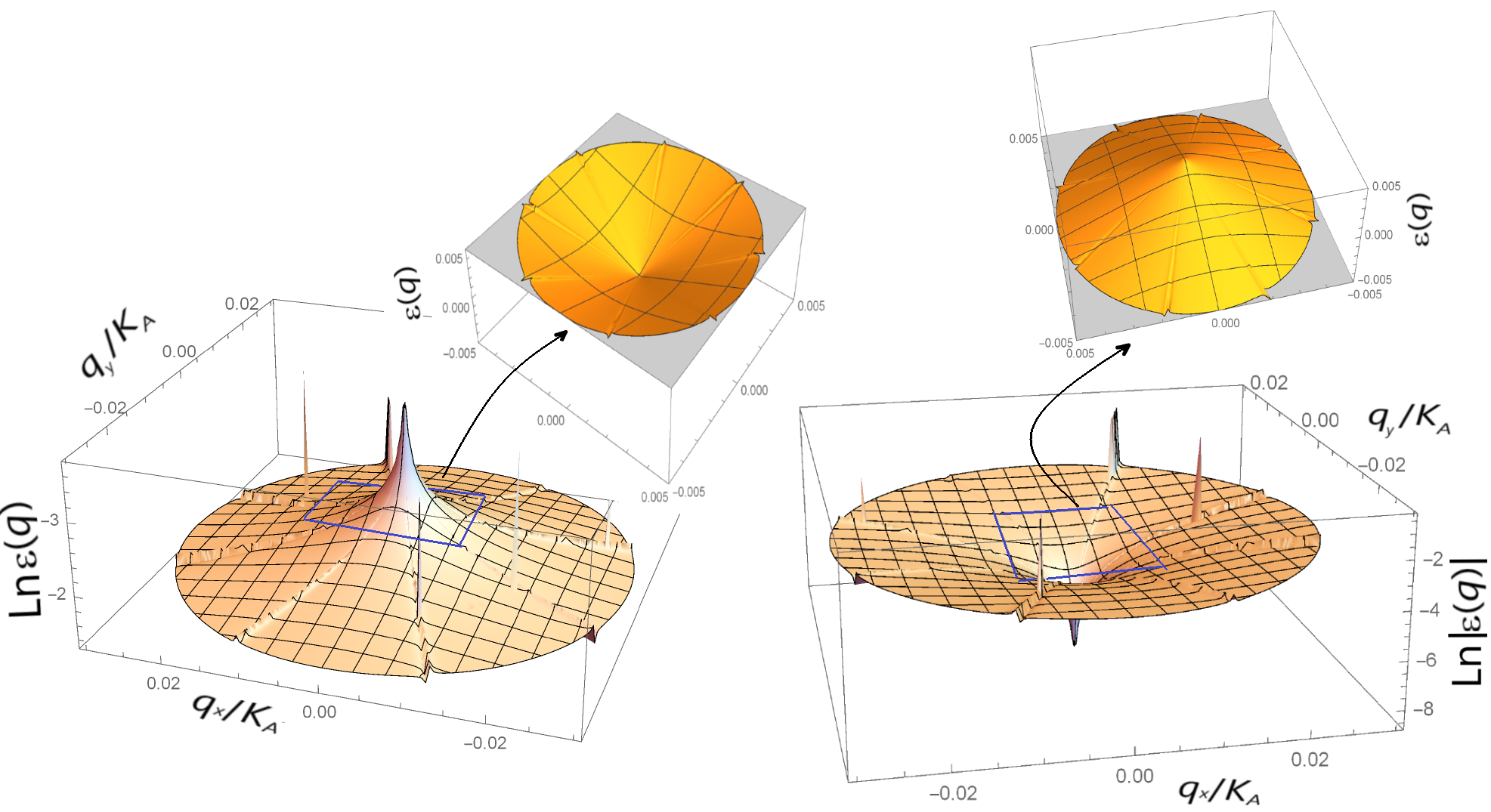} %%% {fig-1_bw.eps for black and white)
\hspace{0.5cm}\includegraphics[width=5.0cm]{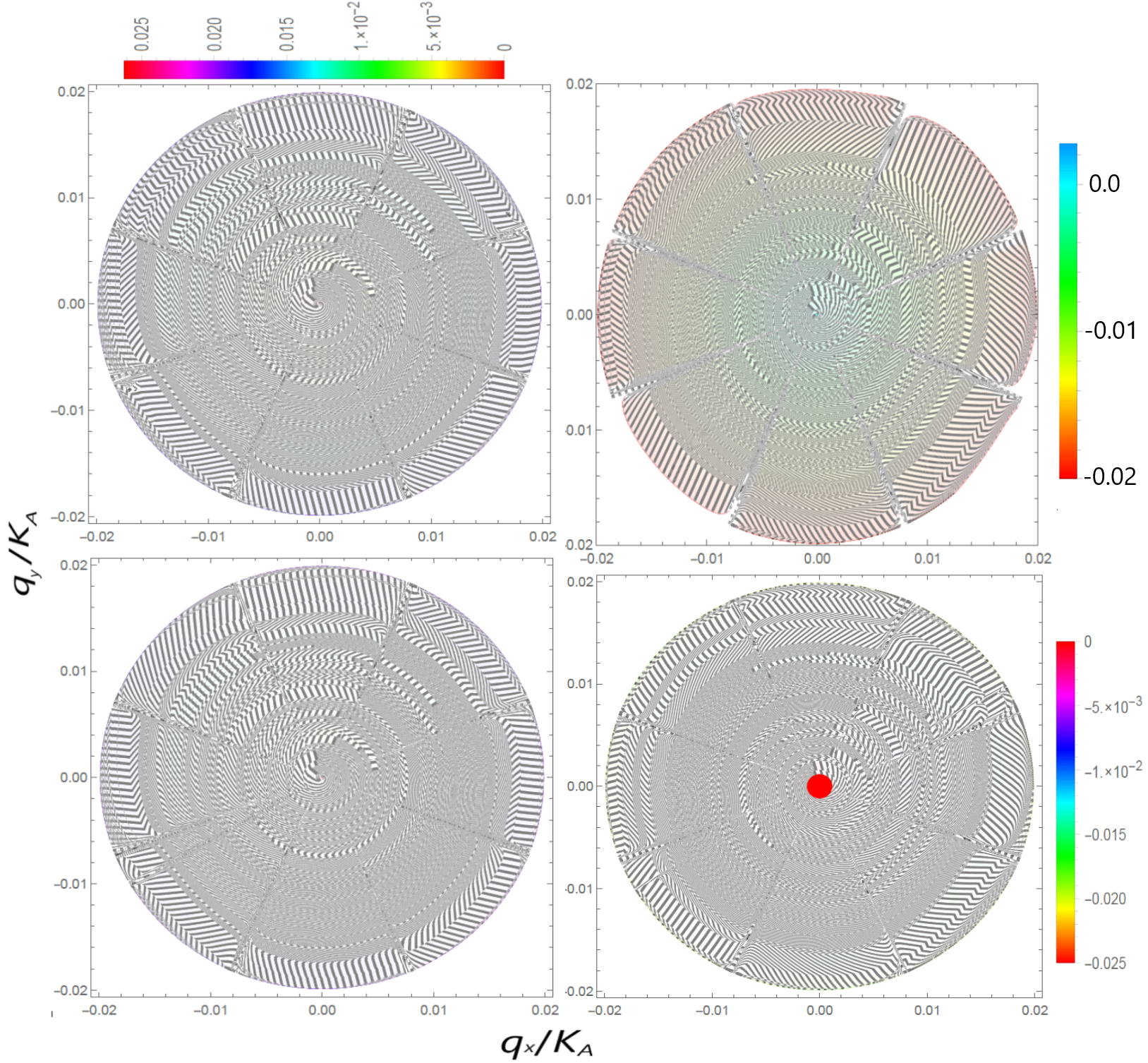} \\
\hspace*{3cm}{(b)}\hspace{5cm}{(d)}\\
\hspace*{0.5cm}\includegraphics[width=7.0cm]{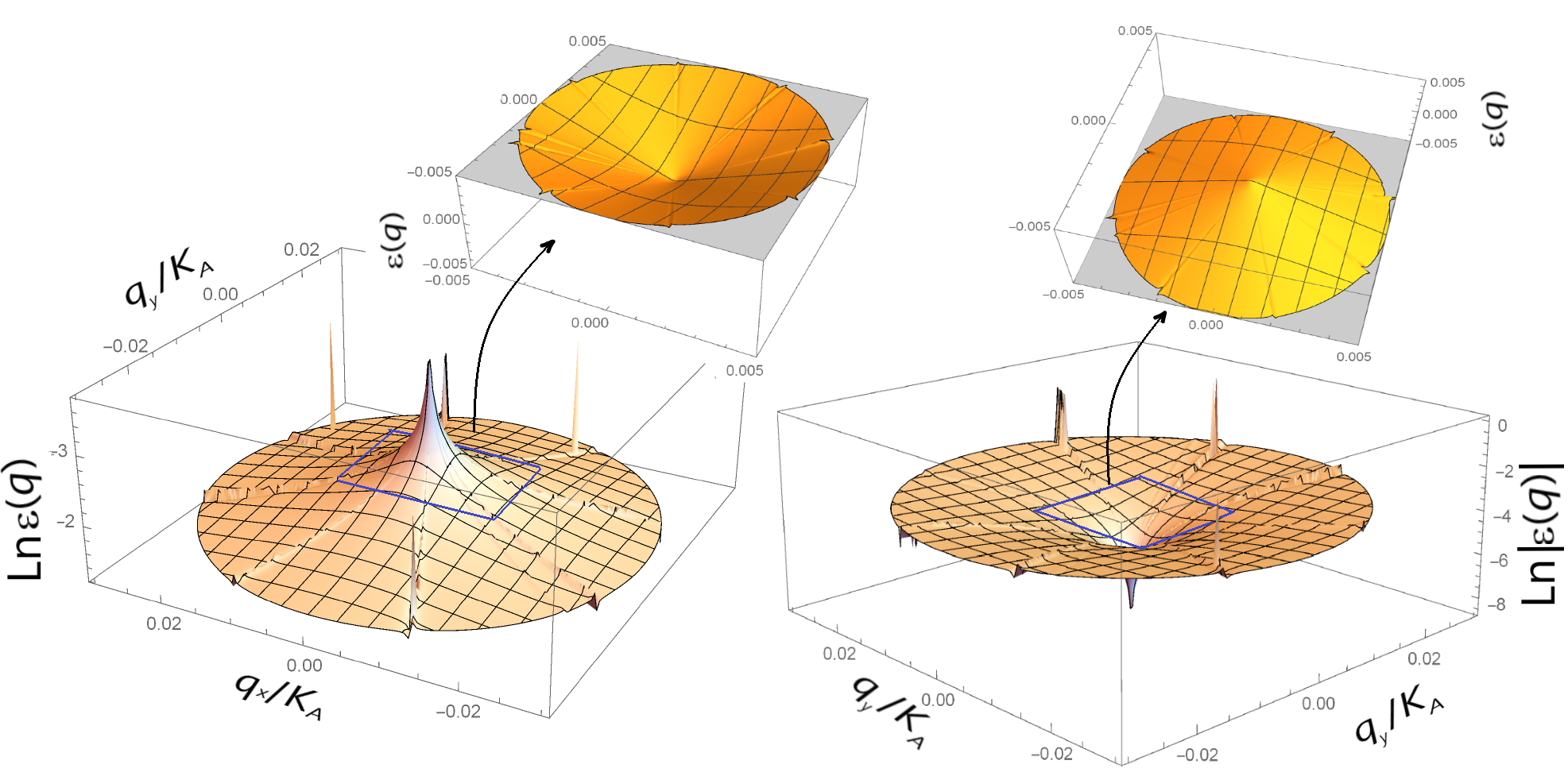} %%% {fig-1_bw.eps for black and white)
\hspace{0.5cm}\includegraphics[width=5.0cm]{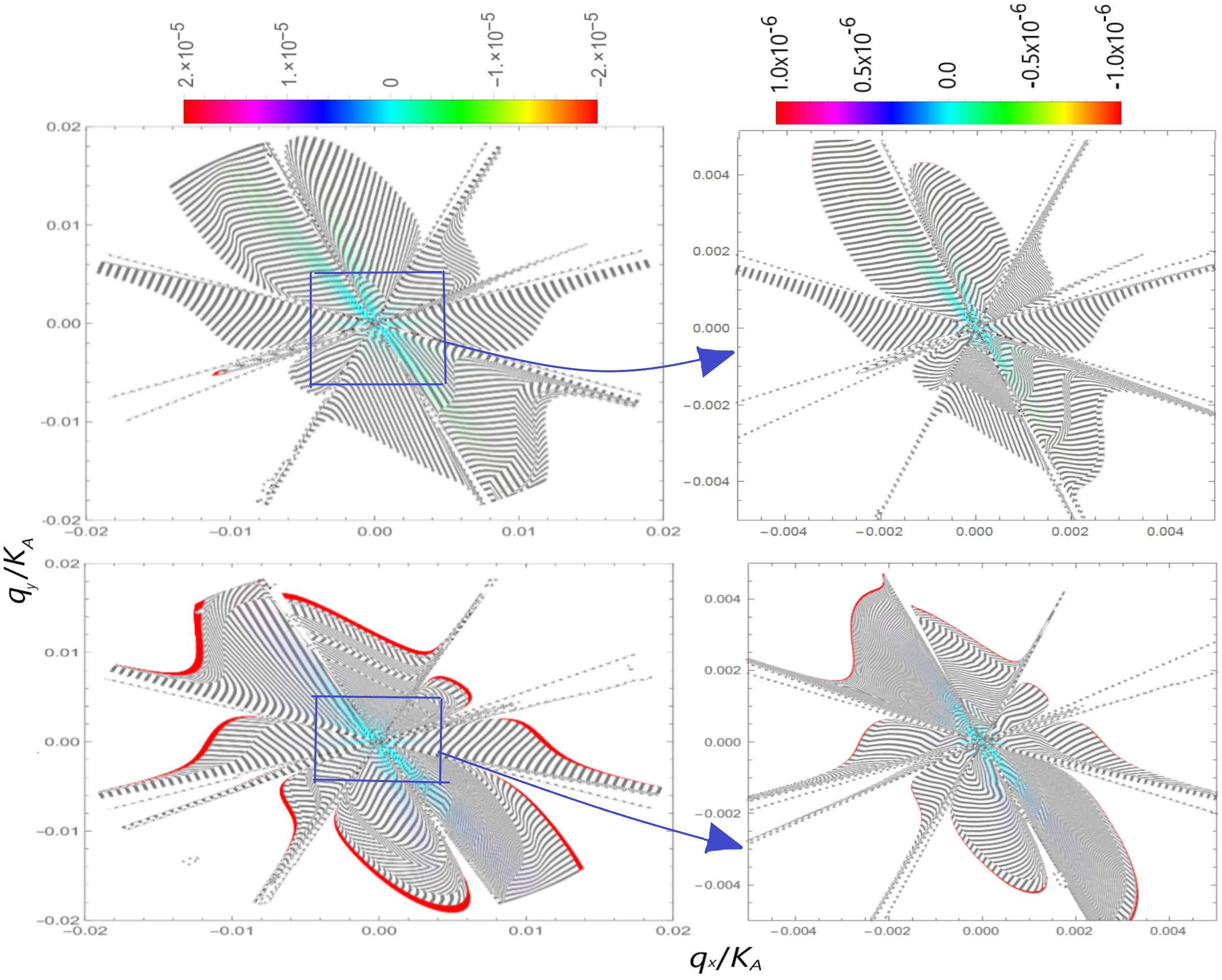}
\caption{Band structure with account of
%с учетом
(a) and  without account of
%без учета
(b) interaction in electron-hole pair
%взаимодействия в
%электрон-дырочной
%паре
on momentum scale %(масштаб):
$q/|K|\sim 0.02$ for quasi-relativistic model of graphene with non-zero
gauge field. (c)  Vortex textures  in contour plots of electron (left) and hole (right) bands
at the non-zero gauge field,
 bands have been calculated
 %рассчитывались
 on momentum scales %(масштаб):
$q/|K|\sim 0.01$ for non-interacting
%для невзаимодействующих
(up) and interacting
%и взаимодействующих
(down) electron and hole constituted
%электрона и дырки,
%составляющих
a Majorana particle. (d) Mass correction to the energy
%Массовая
%поправка к энергетическим
hole (up) and electron (down)
 bands.
 } \label{fig3}
%\end{center}
\end{figure}

%Возрастание спин-орбитального взаимодействия at large $q_y$
%приводит к снятию вырождения Majorana and antiMajorana vortex
%состояний по $J_y$ и, как следствие, к снятию вырождения of the
%subreplicas.
%
Increasing spin-orbit interaction at high $ q_y $
splits the degeneracy of Majorana and antiMajorana
vortex states on $ J_y $ and, correspondingly,  the degeneracy of the subreplicas.
The appearance %Появление
of eight subreplicas is showed in fig.~{fig3}a,b.
%
%This восьмеричное расщепление конусообразных bands  представляет собой a phenomena of Majorana particles deconfinement.
This octal splitting of conical bands represents a phenomenon of Majorana particles deconfinement.
The deconfinement %at high energies
violates the hexagonal symmetry that the high-energetic defects deform the cyclic group $ Z_{12} $
with the generator $ - \pi/6 $ to $ Z_8 $ with the generator $ - \pi/4 $.
Zak phase values, multiple to %кратные
$\pi/4$, appear %появляются
at momenta $q_y>0.2|K|$  as it is shown
%как показано
in fig.~\ref{fig2}a.
%The four vortices располагаются  на сторонах гексагона, чтобы не нарушать the hexagonal symmetry. Эта вихревая
%конфигурация эквивалентна системе типа T-shaped trijunction из трех майорановских мод или антимод.
% The cyclic group $Z_8$  как группа симметрии для систем из трех майорановских мод или антимод в виде
%T-shaped trijunctions схематически представлена in fig.~?????.%
% Несмотря на
%deconfinement, из-за закона сохранения топологического заряда  4
%вихря in the T-trijunction всегда сцеплены на больших энергиях,
%but они становятся асимптотически свободны at $q\to 0$  в окрестности of Dirac points.
%
The deformation is accompanied by subsequent
%сопровожается
%последующим
flatting of the bands that the Fermi velocity trends to 0 as
figs~\ref{fig3}a,b demonstrate. Since the Majorana "force"\ (\ref{variational-Majorana-bispinor})
diverges for the flat bands, four vortices in the T-trijunction are always linked at
high energies, but they become asymptotically free in the vicinity of the Dirac points ($ q \to 0 $).
It testifies that  the conservation law of topological charge    holds.
%
%The deconfinement at the large energies нарушает the hexagonal symmetry  that  the high-energetic
%defects деформируют the cyclic group $Z_{12}$ с образующей $-\pi/6 $ to $Z_8$ с образующей $-\pi/4 $.

%Конфигурация из 4-х deconfined vortices эквивалентна системе типа T-shaped trijunction
%из трех майорановских мод или антимод.
%
A configuration of four deconfined vortices is equivalent to a T-shaped trijunction
system of three Majorana modes or anti-modes.
%
%Как показывает in fig.~???, обход of the T-shaped trijunction по
%контуру c поворотом четыре раза на угол
% $4\pi$ дает набег фазы волновой функции на $2\pi$ в то время, как при обходе одного вихревого дефекта волновая
% функция приобретает фазу $\pi/4$.
 %
 As it is shown in fig.~\ref{fig2}b, the bypass of the T-shaped trijunction over
a contour with four turns by an angle of $ 4 \pi $ gives a phase shift
for the wave function by $ 2\pi $, while at bypass of a single vortex defect
the wave function acquires the phase $ \pi/4 $.
 %

%Расщепление дираковских конусов by strong spin-orbital coupling на
%несовпадающие 4 электронные и 4 hole subreplicas implies нарушение
%электрон-дырочную симметрию.
%
The splitting of the Dirac cones by strong spin-orbital coupling into
non-coinciding four electron and four hole subreplicas implies the break of electron-hole symmetry.
%
%Это нарушение выявляется в виде
%появления добавочных topological Zak фаз equal to $\pi/2, \pi$,
%and correspondingly a additional cyclic group $Z_4 $  at  значениях
%of momentum $q_y>0.24|K|$.
%
This violation is revealed  as an appearance of additional topological Zak phases equal to $ \pi/2, \pi $, and
correspondingly an additional cyclic group $ Z_4 $ at values of momentum $ q_y> 0.24 | K | $.
%Неинвариантные относительно изменения знака энергетической зоны $E$ вида  $E\to -E$ резонансы и
%антирезонансы наблюдаются в the band structure (see fig.~\ref{fig3}a,b) как
%проявление нарушенной электрон-дырочной симметрии by the strong spin-orbital coupling.
%
Resonances and antiresonances that are non-invariant with respect
to change in  sign of the energy band $ E $ of the form
$ E \to -E $ are observed in the band structure (see fig.~\ref{fig3}a,b)
as a manifestation of broken electron-hole symmetry due to strong spin-orbital coupling.
%
% Вихри рождаются парами.
 Vortices are created in pairs. % Therefore, можно
%предположить that the observed резонансы and anti-resonances are
%respectively коры (стоки) and anti-core (источники) вихрей,
%оставшихся после разрушения их пары at the spin-orbital coupling.
%
Therefore, one can assume that the observed resonances and anti-resonances
are respectively the cores (sinks) and anti-core (sources) of the vortices
remaining after the destruction of their pair at the spin-orbital coupling.
%
%Такие топологические дефекты являются монополями и называются
%вейлевскими нодами and antinodes.
%
Such topological defects are monopoles and are called Weyl nodes and antinodes.
%
%Так как группа гомотопии $Z_4 $, то the Weyl nodes (antinodes) in the quasirelativistic graphene
%model удвоены.%
Since the homotopy group is $ Z_4 $, the Weyl nodes (antinodes) in the quasirelativistic graphene model are doubled.
%
%Поскольку спиновые состояния являются крамеровскими
%дуплетами, то это удвоение объясняется нарушением симметрии
%обращения времени, снимающем крамеровское вырождение по спину
%вейлевских нод и антинод.
%
Since spin states are Kramers doublets, this doubling can be explained
by violation of the symmetry of time reversal, that splits the Kramers degeneracy
on spin of Weyl nodes and antinodes.
%
%Мы также изучили влияние взамодействия между электронами и holes в
%парах.
We have also investigated the effect of the interaction between
electrons and holes in pairs that the Majorana mass operator
$M_{AB}$ ($M_{BA}$) is a dynamical mass due to this interaction.
It turns out that this interaction preserves the vortex structure
both Majorana %Оказалось, что это взаимодействие сохраняет,
% вихревую структуру и майорановского, и
and anti Majorana excitation in the vicinity
%в окрестности
of the Dirac points (see figs.~\ref{fig2}c, \ref{fig3}c). The Majorana
mass correction to energy %массовая поправка к энергетическим
hole (up) and electron (down) bands, presented
%представленная
in fig.~\ref{fig3}d, is very small, of the order of
%порядка
$10^{-5}, 10^{-6}$ and is vanishing
%принимает нулевое значение
in $K (K')$. Beyond the Dirac points the degree of chirality
%За
%пределами дираковских точек степень киральноcти
for one of two
%одного из двух
Majorana excitations (either Majorana or
%либо майорановское, либо
anti Majorana excitation) changes that the vortex picture in
the form of distribution of numerous vortex sleeves
changes for only one of the eigenvalues of the operator of dynamical
mass
%% изменяется,
%так как вихревая картина в виде распределения множества рукавов
%вихря меняется для одного из собственных значений оператора
%динамической массы
$M_{AB}$ ($M_{BA}$)
% only, согласно сравнения
in accord to comparison of the vortex structures calculated with
accounting of and without accounting of the interaction in
electron hole pair and presented in a contour plots
% вихревых
%структур, рассчитанных с учетом и без учета взаимодействия в
%электрон-дырочной паре и представленных на контурных графиках
of
figs.~\ref{fig2}c and \ref{fig3}c,d.

Thus, chiral symmetry
%киральная симметрия
of the quasirelativistic graphene model exists not only
%существует
%не только
in the Dirac point, but beyond it. As fig.~\ref{fig3}d demonstrates, the mass term leads to the appearance of the center of
inversion in band structure. The inversion symmetry excludes the mirror symmetry that
the Dirac point is split on the Weyl node and antinode.
Besides, in accordance with the comparison of the band structures calculated with accounting and without accounting of
interaction in the electron-hole pairs
%Кроме того, согдасно сравнения
%вихревых структур, рассчитанных с учетом и без учета
%взаимодействия в электрон-дырочной паре,
in figs.~\ref{fig3}a,b  the shift of Weyl nodes
%сдвиг
%вейлевских нод
and antinodes into the region of higher energies
%в область больших
%энергий
testifies that
the mass-term shifts the location of  Weyl nodes
and antinodes into the region of higher energies and accordingly the  Majorana-like modes can exist
without mixing with the nodes.

\section{Conclusion}
Let us summarize our findings.
Simulations of
non-Abelian Zak phases and band structure of the quasi-relativistic graphene model have been
performed within two approximations of zero- and
non-zero values of gauge  fields.
It has been found that in the approximation of zero gauge field, Zak phases discrete set
testifies that the Dirac point is topologically trivial one. Outside of the valleys the set is
the cyclic group $\mathbf{Z}_2\times\mathbf{Z}_4$. This cyclic group testifies that
the electron-hole degeneration is broken
leading to emergence of double Weyl nodes and antinodes.
The approximation of non-zero gauge fields turns out to be topologically non-trivial
at all energy scale. In this case the Zak phase set forms a cyclic group $\mathbf{Z}_{12}$
which at sufficiently high momenta is deformed into $\mathbf{Z}_8$.
It implies that topological vortex defects
 observed in contour plots of graphene band structure
  are confined by hexagonal symmetry in the Dirac point.
Increasing spin-orbit interaction at high momenta leads to deconfinement
of Majorana modes. The deconfinement
 violates the hexagonal symmetry that the high-energetic defects deform the cyclic group
$\mathbf{Z}_{12}$  to $\mathbf{Z}_8$.
The following effects of the Majorana  mass term  on  the graphene band
structure have been discovered. The first one is the emergence of inverse symmetry
at "switching on"\ of the mass term. The second one is a mass-term shifting of  Weyl nodes
and antinodes into the region of higher momenta. It signifies that after the deconfinement  pure Majorana-like
state can exist in sufficiently large energy region.

\end{document}